\documentclass[10pt,a4paper]{article}
\usepackage[latin1]{inputenc}
\usepackage[T1]{fontenc}
\usepackage[sort]{natbib}
\bibpunct{[}{]}{,}{n}{}{,}

\usepackage{array}
\usepackage{dcolumn}
\newcolumntype{d}[1]{D{.}{.}{#1}}
\newcommand{\PreserveBackslash}[1]{\let\temp=\\#1\let\\=\temp}

\usepackage{amsmath}
\usepackage{amsfonts}

\usepackage{pstricks}
\usepackage[dvips]{graphicx}
\usepackage[dvips]{color}
\usepackage[figuresright]{rotating}
\newcommand{\chem}[1]   {\ensuremath{\mathrm{#1}}}
\newcommand{\lra}       {\longrightarrow}
\newcommand{\molar}     {\scalebox{0.833}{\rm M}}
\newcommand{\molarsf}   {\scalebox{0.833}{\sf M}}

\newcommand{\hp}        {\chem{H^+}}
\newcommand{\bromat}    {\chem{BrO_3^-}}
\newcommand{\bromid}    {\chem{Br^-}}
\newcommand{\hbro}      {\chem{HBrO}}
\newcommand{\cef}       {\chem{Ce^{4+}}}

\newcommand{\brsyr}     {\chem{HBrO_2}}

\newcommand{\brom}      {\chem{Br^-}}

\newcommand{\ma}        {\chem{MA}}
\newcommand{\brma}      {\chem{BrMA}}

\usepackage{pifont}

\def\UNI-{UNI\/{\textbullet\nobreak}}
\newcommand{\unic}{\textsf{UNI\hspace{-0.5ex}
    \raisebox{0.15ex}{\scalebox{1.0}{$\bullet$}}\hspace{0.1ex}C}}

\DeclareMathOperator{\re}  {Re}
\DeclareMathOperator{\im}  {Im}

\DeclareMathOperator{\I}   {i}
\DeclareMathOperator{\mdot}{\mspace{-2.0mu} \cdot \mspace{-2.0mu}}

\newcommand{\gra}{g_0^{\mathrm{r}}}
\newcommand{\gia}{g_0^{\mathrm{i}}}

\newcommand{\grb}{g_1^{\mathrm{r}}}
\newcommand{\gib}{g_1^{\mathrm{i}}}
\newcommand{\drb}{d_w^{\mathrm{r}}}
\newcommand{\dib}{d_w^{\mathrm{i}}}
\newcommand{\sgr}{\sigma_1^{\mathrm{r}}}
\newcommand{\sgi}{\sigma_1^{\mathrm{i}}}
\newcommand{\ash}{\alpha_\mathrm{s}}

\newcommand{\vua}{\vec{u}}
\newcommand{\vub}{\cc{\vec{u}}}
\newcommand{\vuc}{\vec{v}}

\newcommand{\R}{\mathbb R}

\newcommand{\rin}[2]{#1^\mathrm{#2}}
\newcommand{\E}{\mathrm{e}}
\newcommand{\Exp}{\E^{\vec{J}\theta}}
\newcommand{\Exl}{\E^{\vec{p} \mdot \bla \theta}}
\newcommand{\Exi}{\E^{(\vec{p} \mdot \bla - \lambda_i)\theta}}

\newcommand{\Ec}{E^\mathrm{c}}
\newcommand{\Wc}{W^\mathrm{c}}

\newcommand{\bmu}{\boldsymbol{\mu}}

\newcommand{\bdl}{\boldsymbol{\delta}}
\newcommand{\bla}{\boldsymbol{\lambda}}

\newcommand{\BPhi}{\boldsymbol{\Phi}}

\newcommand{\cc}   [1]{{\overline{#1}}}
\newcommand{\stat} [1]{{#1_\mathrm{s}}}
\newcommand{\statb}[1]{{\mathbf{#1}_\mathrm{s}}}

\newcommand{\abs}  [1]{\lvert#1\rvert}

\newcommand{\sci}  [1]{\times 10^{#1}}

\newcommand{\stab} [1]{\vec{#1}_\mathrm{s}}
\newcommand{\ome}  [1]{#1\I\omega_0\vec{I}}
\newcommand{\imag} [2]{\;#1\; \I #2}

\newcommand{\nabn}{\nabla^2}

\newcommand{\diff}[2]{\frac{\partial #1}{\partial #2}}
\newcommand{\difb}[2]{\frac{\partial \vec{#1}}{\partial #2}}

\newcommand{\Dfxx} {\vec{F}_{\vec{x}\vec{x}}}
\newcommand{\Dfxxx}{\vec{F}_{\vec{x}\vec{x}\vec{x}}}
\newcommand{\Dfp}  {\vec{F}_{\mu}}
\newcommand{\Dfxp} {\vec{F}_{\vec{x}\mu}}

\renewcommand{\vec}[1]{\mathbf{#1}}

\newlength{\mlen}

\newcommand{\eno}  [1]{\mbox{\!(\ref{#1})}}
\newcommand{\eqn}  [1]{\mbox{Eq.\! (\ref{#1})}}
\newcommand{\eqns} [2]{\mbox{Eqs.\! (\ref{#1})~and~(\ref{#2})}}
\newcommand{\eqnto}[2]{\mbox{Eqs.\! (\ref{#1})--(\ref{#2})}}
\newcommand{\Eqn}  [1]{\mbox{Equation (\ref{#1})}}

\newcommand{\fig}  [1]{\mbox{Fig.~\ref{#1}}}
\newcommand{\figs} [2]{\mbox{Figs.~\ref{#1}~and~\ref{#2}}}

\newcommand{\figa} [2]{\mbox{Fig.~\ref{#1}#2}}

\newcommand{\tab}  [1]{\mbox{Table~\ref{#1}}}
\newcommand{\tabs} [2]{\mbox{Tables~\ref{#1}~and~\ref{#2}}}

\newcommand{\sect} [1]{\mbox{Section~\ref{#1}}}
\newcommand{\sects}[2]{\mbox{Sections~\ref{#1}~and~\ref{#2}}}

\newcommand{\app} [1]{\mbox{Appendix~\ref{#1}}}

\makeatletter
\def\@eqnacr{{\ifnum0=`}\fi\@ifstar{\@yeqnacr}{\@yeqnacr}}
 
\def\@yeqnacr{\@ifnextchar [{\@xeqnacr}{\@xeqnacr[\z@]}}
 
\def\@xeqnacr[#1]{\ifnum0=`{\fi}\cr \noalign{\vskip\jot\vskip #1\relax}}
 
\def\eqalign{\null\,\vcenter\bgroup\openup1\jot \m@th \let\\=\@eqnacr
\ialign\bgroup\strut
\hfil$\displaystyle{##}$&$\displaystyle{{}##}$\hfil\crcr}
\def\endeqalign{\crcr\egroup\egroup\,}
\makeatother

\usepackage{calc}
\newlength{\Xsize}
\newlength{\Ysize}
\newlength{\newXsize}
\newlength{\newYsize}
\newsavebox{\FigSave}

\newcommand{\ScaleXSize}[1]{
  \sbox{\FigSave}{#1}
  \settowidth{\Xsize}{\usebox{\FigSave}}
  \settoheight{\Ysize}{\usebox{\FigSave}}
  \setlength{\newXsize}{\Xsize*\ratio{\textwidth}{\Xsize}}
  \setlength{\newYsize}{\Ysize*\ratio{\textwidth}{\Xsize}}
  \resizebox{\newXsize}{\newYsize}{\usebox{\FigSave}}
  }

\newcommand{\CircBox}[1]{%
  \pscirclebox[fillstyle=solid, fillcolor=lightgray, framesep=1.0pt, linewidth=0.5pt]{#1}
}

\newcommand{\BlckBox}{\pscircle[fillstyle=solid, fillcolor=black](0,0){1.0mm}}

\begin{document}

\title{Amplitude Equations for Reaction-Diffusion Systems with
  a Hopf Bifurcation and Slow Real Modes}

\author{M.\ Ipsen\footnotemark[2]\
   \and F.\ Hynne\footnotemark[3]\
   \and P.\ G.\ Sørensen\footnotemark[3]%
}

\renewcommand{\thefootnote}{\fnsymbol{footnote}}
\footnotetext[2]{%
  \unic, 
  Danish Computing Center for Research and Education,
  The Technical University of Denmark, 
  Building~304,
  DK-2800 \mbox{Lyngby},
  Denmark, (corresponding author, email: \texttt{mads.ipsen@uni-c.dk}).
  }
\footnotetext[3]{%
  Department of Chemistry,
  University of Copenhagen,
  H.C.Ørsted Institute,
  Universi\-tets\-parken~5,
  DK-2100 Copenhagen,
  Denmark.
  }
\renewcommand{\thefootnote}{\arabic{footnote}}
\maketitle

\begin{abstract} 
  Using a normal form approach described in a previous paper
  we derive an amplitude equation for a reaction-diffusion
  system with a Hopf bifurcation coupled to one or more slow
  real eigenmodes.  The new equation is useful even for
  systems where the actual bifurcation underlying the
  description cannot be realized, which is typical of chemical
  systems.  For a fold-Hopf bifurcation, the equation
  successfully handles actual chemical reactions where the
  complex Ginzburg-Landau equation fails.  For a realistic
  chemical model of the Belousov-Zhabotinsky reaction, we
  compare solutions to the reaction-diffusion equation with
  the approximations by the complex Ginzburg-Landau equation
  and the new distributed fold-Hopf equation.

  \vspace{2.5mm}%

  \vspace{2.5mm}%
\end{abstract}
\vspace*{\fill}

\newpage
\section{Introduction}
\label{sec:Intro}
Near a bifurcation, the evolution of a dynamical system
exhibits critical slowing down, which often admits a
simplified description in terms of an amplitude equation.  We
demonstrate in this paper that slow modes may sometimes be
described in terms of amplitude equations even if the
underlying bifurcation cannot be realized for a given system.
More specifically, we show that slow modes additional to the
critical modes of a bifurcation may be treated in terms of a
higher-codimension bifurcation.

Near a Hopf bifurcation, a reaction-diffusion system may
sometimes be described by a complex Ginzburg-Landau equation
(CGLE\footnote{In this paper, we use the following
abbreviations: CGLE: ``Complex Ginzburg-Landau equation'';
DSHE: ``Distributed slow-Hopf equation''; BZ:
``Belousov-Zhabotinsky''.}), where the motion is restricted to
the unfolded center manifold of the focus.  Diffusion may tend
to move the system off that slow manifold, but fast
non-oscillatory modes may keep the state point close to it.
However, if the system has a slow real mode, diffusion may
significantly excite that mode, thus making a description in
terms of the CGLE dubious.

In principle, a system with a Hopf bifurcation and a slow real
mode may always be treated by a CGLE provided we work
sufficiently close to the bifurcation point.  But experimental
limitations may prevent us from operating close enough.  Slow
modes are likely to appear in complex chemical or biochemical
reactions where the spectrum of characteristic times often
spans many orders of magnitude.  An example is the
Belousov-Zhabotinsky (BZ) reaction which under certain
conditions shows simple periodic oscillations.  These basic
oscillations may be described by relatively simple models like
the Oregonator~\citep{OregOrig}.  At other operating points,
the BZ reaction may exhibit complex oscillations (including
chaos)~\citep{ZhabOverview,Wang95}, which can be attributed to
the presence of a slow real mode and cannot be described by
the Oregonator.  But the complexity of the BZ reaction can be
described quite well by a four-dimensional extension of the
Oregonator model~\citep{Wang95}.  For that more realistic
model, we show in this paper that a description by the CGLE
fails at experimentally feasible conditions.

Treating a reaction-diffusion system by an amplitude equation
saves computational resources and provides interpretations
that give more insight into the character of the solution.  It
is therefore highly desirable to develop an amplitude
description to replace the CGLE for important chemical systems
like the BZ reaction or biochemical reactions such as
glycolysis.  In this paper, we shall derive an amplitude
equation for reaction-diffusion systems with a Hopf
bifurcation and a slow real mode.  The equation is based on
the amplitude equation for a fold-Hopf bifurcation even though
the system need not posses such codimension-two bifurcation.
We shall refer to the result as a distributed slow-Hopf
equation (DSHE).  We show that the four-dimensional
reaction-diffusion model of the BZ reaction may be excellently
described in terms of the appropriate DSHE.

The development of the DSHE involves several intermediate
results that are important in their own right.  We first
derive a set of amplitude equations for the class of
bifurcations (occurring in the corresponding spatially
homogeneous system) with one pair of pure imaginary
eigenvalues and any number of semisimple eigenvalues zero.
(Chemical and biochemical reactions may well have several slow
modes.)

The method used for deriving the amplitude equations builds on
results previously obtained for ordinary differential
equations~\citep{Ipsen98}.  The unfolded center manifold for
the corresponding homogeneous system is parameterized by
points $\vec{y}$ in the center subspace.  At each point of
(physical) space, the state point is restrained to move on the
center manifold, and the amplitude equation is found as the
equation for the modulation of the oscillations of $\vec{y}$
(described in a basis of critical eigenvectors) obtained from
the restrained reaction-diffusion equation.

The resulting class of amplitude equations for the general
case is given by \eqn{eq:FinalAmp} with
\eqns{eq:DiffDef}{eq:DiffDefTwo}.  From a solution to an
amplitude equation, we get the physical solution using the
transformation~\eno{eq:HFinal}, determined as indicated in
Appendix A (see also~\citep{Ipsen98}).

We treat the special case of a distributed fold-Hopf
bifurcation in detail in \sect{sec:FoldHopfDer} (see
\eqn{eq:FoldHopf} together with \tab{tab:FoldHopfTable}).  We
note also that the CGLE is another special case; thus the
present work contains a new derivation of that important
equation.

We modify the distributed fold Hopf equation to get an
amplitude equation, the DSHE, applicable to systems with a
Hopf bifurcation and one slow but non-critical real mode.  The
result is~\eqn{eq:DSHE}, which can be scaled to the
dimensionless form \eqn{eq:ScaledFold}.  The precise relation
of the DSHE to the original CGLE is required when the two
equations are compared.  It is explained in \app{sec:CGLERel}.

The approximations in terms of various amplitude equations are
evaluated for a four-dimensional extension of the Oregonator
model mentioned above, which is described in
\sect{sec:BZModel}.  The model provides a realistic
description of the BZ reaction based on extensive kinetic
work~\citep{FKNorig,FFrate} and describes much of the
dynamical behavior of the BZ system qualitatively or
semiquantitatively correctly (see~\citep{Wang95} for a
description of the complex dynamics and \citep{tina2} for the
bare oscillations).

We compare the ``exact'' solution obtained by direct numerical
integration of the reaction-diffusion equation with the
approximations obtained as solutions to the amplitude
equations, the CGLE on the one hand and the DSHE on the other.
We find in \sects{sec:CGLEDes}{sec:DSHEDes} that whereas the
simple CGLE fails, the DSHE provides a very accurate
description of the dynamics of the reaction-diffusion system.

The result shows the effect of a slow real mode near a Hopf
bifurcation, and that a Ginzburg-Landau approximation cannot
be considered reliable for the BZ system.  The slow real mode
is a natural (unavoidable) part of that system.  A similar
effect has also been considered by Aranson et.\ al
\citep{Ar96} who coupled the FitzHugh-Nagumo model
\citep{Fitzhugh61,Nagumo62} with an artificial slow real mode.
There the slow mode interacted with an excitable system, and
the effect was the opposite of the one reported here for the
oscillatory BZ system.  Random initial conditions normally
give rise to a disordered distribution of spiral waves in the
FitzHugh-Nagumo model, but the slow mode resulted in the
development of a single spiral covering the entire spatial
domain.

\section{Derivation of amplitude equations}
\label{sec:Derivation}
In this section, we derive amplitude equations for chemical
reaction-diffusion systems for which the (spatially)
homogeneous system is near a bifurcation at which the spectrum
of critical eigenvalues consists of one pair of complex
conjugate, imaginary eigenvalues and several real semisimple
eigenvalues equal to zero.

\subsection{Homogeneous systems}
\label{sec:HomoSyst}
We consider a dynamical system, described by a vector $\vec{x}
\in \R^n$, and depending on a set of parameters $\bmu \in
\R^s$.  The system is supposed to have a stationary point
$\stab{x}(\bmu)$ at which a local bifurcation occurs at $\bmu
= \vec{0}$.  We shall use the stationary point,
$\stab{x}(\vec{0})$, at the bifurcation point as origin.  The
homogeneous chemical system evolves according to the kinetic
equation

\begin{equation}
  \frac{d\vec{x}}{dt} = 
  \vec{F}(\vec{x};\bmu)  = 
  \vec{J} \mdot \vec{x} + \vec{f}(\vec{x},\bmu).
  \label{eq:ODE}
\end{equation}
The Jacobian matrix at the stationary point at the
bifurcation, $\vec{J} = D\vec{F}(\vec{0},\vec{0})$, has $r$
semisimple eigenvalues $\lambda_i$ and $r$ pairs of linearly
independent right and left eigenvectors $\vec{u}_i$ and
$\vec{u}^*_i$, normalized according to the biorthonormality
relations
\begin{equation}
  \vec{u}_i^* \mdot \vec{u}_j = \delta_{ij}, 
  \quad \text{for \ }
  i,j = 1,\dots,r.
  \label{eq:BiOrtho}
\end{equation}

The $r$-dimensional center manifold $\Wc$ at $\bmu = \vec{0}$
is tangent to the center subspace $\Ec$ at $\vec{x} =
\vec{0}$, which is spanned by the $r$ right eigenvectors
$\vec{u}_i$.  Near $\vec{x} = \vec{0}$, motion in $W^c$ (or
its unfolding) is usually much slower than motion towards it
or away from it (critical slowing down), and the motion
outside $W^c$ may often be considered transient and fast and
therefore disregarded.  This fact adds to the practical
importance of the center manifold.

We have previously \citep{Ipsen98} derived simple, explicit
expressions that allow us to write down the normal form
equation for the flow on the center manifold and unfoldings of
it (for $\bmu \ne \vec{0}$) as well as a transformation from
normal form coordinates to the original physical coordinates.
The transformation has the form of a map
\begin{equation}
  \vec{x} = \vec{y} + \vec{h}(\vec{y},\bmu), \quad
  \text{where}\; \vec{y} \in \Ec,
  \label{eq:NormalTrans}
\end{equation}
from $\Ec \times \R^s$ to the (unfolded) center manifold. For
oscillatory modes, the normal form allows a formulation in
terms of ``amplitudes'' by a straightforward modification of
the normal form equation.  Consequently, we sometimes refer to
the equation as an ``amplitude equation'', whether it is
modified or not.  The results of the previous work needed in
the present paper are summarized in~\app{app:AmpTrans}.

From the results of \app{app:AmpTrans}, we may easily obtain
an explicit expression for the center manifold, parameterized
by coordinates in the center subspace in a basis of critical
eigenvectors, valid for any given local bifurcation (with
semisimple critical eigenvalues).  We refer to such
representation as an ``amplitude transformation''.  It is
given by \eqn{eq:NormalTrans} with $\vec{h}$ expanded in
powers of the coordinates of $\vec{y}$, see
\eqn{eq:AppCenterExp} of \app{app:AmpTrans}.  Observe that
this expansion also provides an explicit expression for the
vector field on $\Wc$.  We shall make use of these two results
in the next subsection to derive an amplitude equation for the
corresponding reaction-diffusion problem.

\subsection{Extension to spatially inhomogeneous systems}
\label{sec:ExtST}
When the system described by \eqn{eq:ODE} is modified by
introducing diffusion of the involved physical quantities, the
time evolution of the system is governed by a
reaction-diffusion equation
\begin{equation}
  \difb{x}{t} = 
  \vec{F}(\vec{x};\bmu) + \vec{D} \mdot \nabn \vec{x},
  \label{eq:ReacDiff}
\end{equation}
where $\vec{x} = \vec{x}(\vec{r},t)$ depends on a spatial
position $\vec{r}$ as well as time $t$, and $\vec{D}$ is a
diffusion operator, which typically, to a good approximation,
can be represented by a diagonal matrix.


In a homogeneous system, the (unfolded) center manifold is
invariant under the flow.  In chemical reaction-diffusion
systems, diffusion may take the local concentration away from
$\Wc$ even if it initially was on $\Wc$ everywhere.

However, close to the bifurcation, the motion on the unfolded
center manifold due to the chemical reactions often is much slower
than motion transverse to it, as already mentioned.  Consequently,
diffusion never takes the system far away from that ``slow
manifold'', so the evolution of the spatial system may be
approximately described by \eqn{eq:ReacDiff} with $\vec{x}$
restrained to move on the slow manifold $\Wc(\bmu)$ given
by \eqn{eq:NormalTrans}.

We may therefore simply substitute the right hand side
of \eqn{eq:NormalTrans} for $\vec{x}$ in \eqn{eq:ReacDiff} to
get a differential equation in $\vec{y}$, the corresponding
point in the center subspace.  The appropriate solution to the
resulting equation is then transformed by \eqn{eq:NormalTrans}
to the motion in the center manifold, the approximate solution
to the reaction-diffusion equation~\eno{eq:ReacDiff}.

Unfortunately, direct substitution of the parameterization for
the slow manifold $\Wc(\bmu)$ in \eqn{eq:ReacDiff} results in
``counter-rotating terms''~\citep{BohrTurb} that prevent
straightforward use of amplitudes for oscillatory modes.  By
averaging, these recalcitrant terms can be seen to be
negligible close enough to the bifurcation point, but that
method is approximate and inelegant.  We shall therefore use a
method of two times while still building on the result of the
theory outlined in \app{app:AmpTrans}.

We now restrict the discussion to a bifurcation with a single
pair of complex conjugate imaginary eigenvalues $\pm\I
\omega_0$ together with any number of semisimple zero
eigenvalues.  There are $r$ such critical eigenvalues
(counting multiplicity) and $r$ linearly independent right
eigenvectors and corresponding left eigenvectors satisfying
the biorthonormality relations~\eno{eq:BiOrtho}.  The period
of the oscillations near $\vec{x} = \vec{0}$ at $\bmu =
\vec{0}$ is $T = 2\pi/\omega_0$.

We express the solution $\vec{y}(\vec{r},t)$ as the modulation
of harmonic oscillations of frequency $\omega_0$ for the
oscillatory mode and seek an equation for the modulation.  To
that end, it is convenient to view $\vec{y}$ as a function of
two independent time variables, $\tau$ and $\theta$, with
$\theta$ accounting for the harmonic oscillation (of the
oscillatory degrees of freedom) and $\tau$ describing the
modulation.

We define 
\begin{equation}
  \vec{y}(\vec{r},\tau,\theta) = 
  \Exp \mdot \vec{z}(\vec{r},\tau) = 
  \sum_{i=1}^r \E^{\lambda_i \theta} z_i(\vec{r},\tau)
  \vec{u}_i,
  \label{eq:YDef}
\end{equation}
where $\tau$ and $\theta$ are given functions of time.  At the end
of the calculation, we shall choose
\begin{equation}
  \tau = \theta = t.
  \label{eq:timedef}
\end{equation}
To obtain the correct equation for
$\vec{y}(\vec{r},\tau,\theta)$, we use \eqn{eq:timedef}
initially as
\begin{equation}
  \frac{\partial \vec{y}}{\partial t} =
  \frac{\partial \vec{y}}{\partial \theta} 
  \frac{d\theta}{dt} +
  \frac{\partial \vec{y}}{\partial \tau} 
  \frac{d\tau}{dt} =
  \frac{\partial \vec{y}}{\partial \theta} +
  \frac{\partial \vec{y}}{\partial \tau}.
  \label{eq:TimeDecmp}
\end{equation}
but otherwise we use the variables $\vec{r}$, $\tau$, and
$\theta$ consistently throughout (so no confusion should arise
from using the symbol $\vec{y}$ for the two different
functional forms).

We now view $\vec{x}$ as a function of $\vec{r}$, $\tau$, and
$\theta$ through \eqn{eq:NormalTrans}, and using
\eqn{eq:TimeDecmp}, the reaction-diffusion equation
\eno{eq:ReacDiff} may then be written as
\begin{equation}
  \diff{\vec{x}}{\tau} = 
  (\vec{J} \mdot \vec{x} - \diff{\vec{x}}{\theta}) +
  \vec{f}(\vec{x},\bmu) +\vec{D} \mdot \nabn \vec{x},
  \label{eq:ReacTimeSplit}
\end{equation}
with
\begin{equation}
  \vec{x}(\vec{r},\tau,\theta) =
  \vec{y}(\vec{r},\tau,\theta) +
  \vec{h}(\vec{y}(\vec{r},\tau,\theta),\bmu).
  \label{eq:TauParam}
\end{equation}

For a given physical (or chemical) problem, the transformation
$\vec{h}$ is a well defined function of $\vec{y}$ and $\bmu$.
So for a given $\bmu$, the second term of \eqn{eq:TauParam}
and its derivatives with respect to $\vec{r}$, $\tau$, and
$\theta$ are completely determined by the function
$\vec{y}(\vec{r},\tau,\theta)$.  Consequently, these terms do
not contain any ``new'' information about the dynamics not
already contained in $\vec{y}(\vec{r}, \tau, \theta)$ and the
transformation $\vec{h}(\vec{y},\bmu)$.

These remarks suggest that we try to eliminate as much as
possible of the nonlinear term of
\begin{equation}
  \diff{\vec{x}}{\tau} =
  \diff{\vec{y}}{\tau} +
  \diff{\vec{h}}{\tau} 
\end{equation}
in \eqn{eq:ReacTimeSplit}.  We observe that
\begin{equation}
  \diff{\vec{y}}{\tau} = \Exp \mdot \diff{\vec{z}}{\tau} =
  \sum_{i=1}^r \E^{\lambda_i\theta}
  \diff{z_i}{\tau} \vec{u}_i
\end{equation}
implying that we may extract an equation for
$\diff{z_i}{\tau}$ by multiplying \eqn{eq:ReacTimeSplit} by
$\E^{-\lambda_i \theta} \vec{u}_i^*$ (from the left) and
averaging the result over a period $T$ of $\theta$.  The
left-hand side of \eqn{eq:ReacTimeSplit} then simply gives
$\diff{z_i}{\tau}$, since
\begin{equation}
  \frac{1}{T}
  \int_0^T 
  \E^{-\lambda_i \theta}
  \vec{u}_i^* \mdot \diff{\vec{h}}{\tau} d\theta = 0.
  \label{eq:Average}
\end{equation}
We prove this property in~\app{app:AverageZero}.  There we
also show that
\begin{equation}
  \frac{1}{T}
  \int_0^T 
  \E^{-\lambda_i \theta}
  \vec{u}_i^* \mdot 
  (\vec{J} \mdot \vec{x} - \diff{\vec{x}}{\theta}) d\theta = 0,
  \label{eq:ParZero}
\end{equation}
implying that the two terms in parentheses disappear from
\eqn{eq:ReacTimeSplit} by the above operation.

As argued in \app{app:AmpTrans}, we may use a power series
expansions of $\vec{x}$ and $\vec{f}(\vec{x},\bmu)$ for $\vec{x}$
on the slow manifold $\Wc(\bmu)$.  Here we first note that
\begin{equation}
  \vec{y}^\vec{p} = \prod_i y_i^{p_i} =
  \exp (\sum_i p_i \lambda_i \theta) \prod_i z_i^{p_i} =
  \Exl \vec{z}^\vec{p}
  \label{eq:YProd}
\end{equation}
in concise notation.

For $\vec{f}(\vec{x},\bmu)$, we get using the
expansion~\eno{eq:FExp}
\begin{align}
  \frac{1}{T}
  \int_0^T 
  \E^{-\lambda_i \theta}
  \vec{u}_i^* \mdot \vec{f}(\vec{y}+\vec{h}(\vec{y},\bmu)) d\theta  &=
  \frac{1}{T}
  \sum_\vec{pq} \vec{u}_i^* \mdot \vec{f}_\vec{pq} 
  \vec{z}^\vec{p} \bmu^\vec{q} 
  \int_0^T \Exi d\theta \nonumber\\
  &= \quad\sideset{}{^i}\sum_\vec{pq} f_\vec{pq}^{(i)}
  \vec{z}^\vec{p} \bmu^\vec{q},
  \label{eq:FieldAverage}
\end{align}
in which $f^{(i)}_\vec{pq} = \vec{u}_i^* \mdot
\vec{f}_\vec{pq}$ is the $i$'th component of the vector
coefficient of the expansion of $\vec{f}(\vec{x},\bmu)$ on
$\Wc(\bmu)$ whereas the sum is taken over all sets
$(\vec{p},\vec{q})$ for which the resonance condition for the
$i$'th component
\begin{equation}
  \vec{p} \mdot \bla = \sum_j p_j \lambda_j = \lambda_i
  \label{eq:ResDiff}
\end{equation}
is satisfied.  All other terms vanish because of the integral
of the exponential in \eqn{eq:FieldAverage}.  Note that
critical eigenvalues are either pure imaginary or zero.  For
clarity, the summation over the resonant terms for the $i$'th
component is marked with a superscript $i$ in the sum in
\eqn{eq:FieldAverage}.  For the diffusion term of
\eqn{eq:ReacTimeSplit}, we similarly get (using the
expansion~\eno{eq:AppCenterExp} from \app{app:AmpTrans})
\begin{align}
  &
  \frac{1}{T}
  \int_0^T 
  \E^{-\lambda_i \theta}
  \vec{u}_i^* \mdot \vec{D} \mdot 
  \nabn (\vec{y}+\vec{h}(\vec{y},\bmu)) d\theta  = \nonumber \\
  & 
  \sum_j D_{ij} \nabn z_j \sum_\vec{pq}\frac{1}{T} 
  \int_0^T \E^{(\lambda_j - \lambda_i)\theta} d\theta +
  \vec{u}_i^* \mdot \vec{D} \mdot 
  \sum_\vec{pq} \vec{h}_\vec{pq} \bmu^\vec{q}\nabn \vec{z}^\vec{p} 
  \int_0^T \Exi d\theta = \nonumber \\
  &
  \sum_j D_{ij} \nabn z_j +
  \sideset{}{^i}\sum_\vec{pq} \vec{u}_i^* \mdot \vec{D} \mdot 
  \vec{h}_\vec{pq} \bmu^\vec{q}\nabn \vec{z}^\vec{p}.
  \label{eq:DiffAverage}
\end{align}
Here
\begin{equation}
  D_{ij} = \vec{u}^*_i \mdot \vec{D} \mdot \vec{u}_j
\end{equation}
and the sum over $j$ is taken over all terms for which
$\lambda_j = \lambda_i$, and the sum over $\vec{p}$ and
$\vec{q}$ includes all resonant terms for the $i$'th component
(and no other).  By using \eqn{eq:Average}, \eqn{eq:ParZero},
\eqn{eq:FieldAverage}, and \eqn{eq:DiffAverage}, we obtain an
amplitude equation for the modulation as a set of coupled
equations for the set of all coefficients $z_i(\vec{r},\tau)$
defined in \eqn{eq:YDef}.
\begin{equation}
  \diff{z_i}{t} =
  \sideset{}{^i}\sum_\vec{pq} \vec{f}_\vec{pq} \vec{z}^\vec{p}\bmu^\vec{q} + 
  \sum_j d_{ij} \nabn z_j +
  \sideset{}{^i}\sum_\vec{p} \mathcal{D}_\vec{p}^{(i)} \nabn \vec{z}^\vec{p}.
  \label{eq:FinalAmp}
\end{equation}
Here
\begin{equation}
  d_{ij} = 
  \begin{cases}
    \vec{u}^*_i \mdot \vec{D} \mdot\, 
    (\vec{u}_j + 
    \sum_{\abs{\vec{q}}>0} \vec{h}_{\bdl_j\vec{q}} \bmu^\vec{q}),
     & \text{$\lambda_j = \lambda_i$}\\
  0, & \text{$\lambda_j \neq \lambda_i$}
  \end{cases}
  \label{eq:DiffDef}
\end{equation}
and
\begin{equation}
  \mathcal{D}_\vec{p}^{(i)} = 
  \begin{cases}
    \vec{u}^*_i \mdot \vec{D} \mdot
    \sum_\vec{q} \vec{h}_\vec{pq} \bmu^\vec{q} 
      & \text{$\abs{\vec{p}}    > 1$}\\
   0, & \text{$\abs{\vec{p}} \leq 1$},
  \end{cases}
  \label{eq:DiffDefTwo}
\end{equation}
in which $\abs{\vec{p}} = \sum_j p_j$.  As before, a
superscript $i$ restricts a summation to resonant terms for
the $i$'th component.  In \eqn{eq:DiffDef}, the index set
$\bdl_j$ has components described by the usual Kronecker
delta, $(\bdl_j)_k = \delta_{jk}$.

In \eqn{eq:FinalAmp}, we have identified $\tau$ with the
``real'' time, $t$, since we no longer need to distinguish
between the two ``formal'' time scales.  From a solution
$z_i(\vec{r},t)$, $i=1,\dots,r$, to \eqn{eq:FinalAmp} we get
an (approximate) solution to the reaction-diffusion
equation~\eno{eq:ReacDiff} as
\begin{subequations}
  \begin{align}
    \vec{x}(\vec{r},t) &= 
    \vec{y}(\vec{r},t) + \vec{h}(\vec{y}(\vec{r},t), \bmu),\\
    \vec{y}(\vec{r},t) &= 
    \sum_{i=1}^r z_i(\vec{r},t)\E^{\lambda_i t}\vec{u}_i
  \end{align}
  \label{eq:HFinal}
\end{subequations}
Here we note that the transformation $\vec{h}$ produces
anharmonic terms in $\theta$ in the plane of oscillations as
well as components off the center subspace $\Ec$.  Of course,
these terms are also modulated (by products of powers of the
amplitudes $z_i$) and are therefore not periodic in $t$.

For a spatially homogeneous system, the diffusion terms
disappear from the amplitude equation~\eno{eq:FinalAmp}, and
the result therefore agrees with the general expression for a
homogeneous system as quoted in \app{app:AmpTrans}.  In
general, the amplitude equation contains diffusion terms that
modify the motion on the slow manifold.  The sum over $j$ in
\eqn{eq:FinalAmp} represents the linear diffusion terms which
include couplings between bifurcating modes corresponding to
the same eigenvalue.  But note in particular that the
oscillatory modes (with eigenvalues $\pm\I\omega_0$) do not
couple to any of the real modes---only the real modes may
couple to each other.

The nonlinear diffusion terms, the last sum in
\eqn{eq:FinalAmp}, arise because the motion is restrained to
the slow manifold; diffusion is linear in $\vec{x}$ but we
describe the dynamics in terms of points $\vec{y} \in \Ec$,
and $\vec{x}$ is clearly not linear in $\vec{y}$.  It is to be
expected, that the linear diffusion terms will dominate over
the nonlinear diffusion, and these higher order terms may
therefore be neglected in first approximation.  Below we show
examples which are excellently described using linear
diffusion alone.  In some cases, the nonlinear diffusion may
be important, however.  For example, this may be the case in
situations where the real part of any of the linear diffusion
coefficients becomes negative---a situation which easily arises
in biochemical systems where the diffusion constants may
differ by several orders of magnitude.

In~\sect{sec:FoldHopfDer}, we shall exhibit the amplitude
equation for a fold-Hopf bifurcation (which has one
oscillatory and one real mode) as a special case of
\eqn{eq:FinalAmp}.  We also note how the well-known complex
Ginzburg-Landau equation (CGLE) appears as a special case of
the result~\eno{eq:FinalAmp}.  Thus the present work contains
a novel derivation of the CGLE.

The fold-Hopf bifurcation will be used to solve the
reaction-diffusion equation for a realistic model of the
Belousov-Zhabotinsky reaction---we shall compare the solution
obtained from the new equation with those of the
reaction-diffusion equation itself and of the corresponding
CGLE\@.  In addition, this example also serves as an
illustration of the practical use of \eqn{eq:FinalAmp}.

\subsection{Distributed Fold-Hopf equation}
\label{sec:FoldHopfDer}
In the following, we shall focus on the special case of
\eqn{eq:FinalAmp}, where there is just a single eigenvalue in
addition to the imaginary pair.  This bifurcation is called a
fold-Hopf bifurcation corresponding to the case where a Hopf
bifurcation coincides with a single non-degenerate real
bifurcation.  The linear diffusion terms contain no coupling
between modes, so the sum over $j$ in \eqn{eq:FinalAmp}
reduces to just one term, $d_{ii} \nabn z_i$ with
\begin{equation}
  d_{ii} = \vec{u}_i^* \mdot \vec{D} 
  \mdot\, (\vec{u}_i + \sum_\vec{q}\vec{h}_{\bdl_i\vec{q}} \bmu^\vec{q}).
\end{equation}
Here we shall exhibit the amplitude equation~\eno{eq:FinalAmp}
to lowest non-trivial order in $\bmu$ (\emph{i.e.} including
only terms essential to the unfolding of local terms) implying
that the diffusion term $d_{ii}$ reduces to $d_{ii} =
\vec{u}_i^* \mdot \vec{D} \mdot \vec{u}_i$.

To lowest order, we need only terms linear in $\bmu$ in
general, so we get similar independent terms from each
component of $\bmu$.  Therefore, we shall exhibit the result
for a single scalar $\mu$ even though one is often interested
in using two independent parameters for bifurcations of
codimension two: the generalization is straightforward.

For convenience, we simplify the notation by assigning
$\lambda_{1,2} = \pm\I\omega_0$ and $\lambda_3 = 0$ for the
eigenvalues at the bifurcation point, and $\vec{u}_1 =
\vec{u}$, $\vec{u}_2 = \cc{\vec{u}}$, and $\vec{u}_3 =
\vec{v}$ for the corresponding right eigenvectors.  In this
case, we need only consider the amplitudes for $w = z_1$ and
$z = z_3$ since complex conjugation gives $z_2 = \cc{w}$.

If we include all terms up to third order in $w$ and $z$ for
$q = 0$ and only terms essential to the unfolding for $q > 0$
and neglect nonlinear diffusion, we then obtain the following
amplitude equation for a fold-Hopf bifurcation in a
reaction-diffusion equation~\eno{eq:ReacDiff}
\begin{subequations}
  \label{eq:FoldHopf}%
  \begin{align}
    \dot{w} &= \sigma_1 \mu w + g_0 w z + g_1 \abs{w}^2w + g_2 w z^2 +
               d_w \nabn w,\\
    \dot{z} &= \rho_0 \mu + c_0 \abs{w}^2 + c_1 z^2 + 
               c_2 \abs{w}^2 z + c_3 z^3 + d_z \nabn z.
  \end{align}
\end{subequations}
Here we have introduced the following compact notation
\begin{align}
  \label{eq:FoldHopfCoeff}
  d_w &= \vec{u}^* \mdot \vec{D} \mdot \vec{u},  &
  d_z &= \vec{v}^* \mdot \vec{D} \mdot \vec{v},  &{}&{} &{}&{} \nonumber\\
  \sigma_1 &= \vec{u}^* \mdot \vec{f}_{1001},      &
    \rho_0 &= \vec{v}^* \mdot \vec{f}_{0001},      &{}&{} &{}&{} \nonumber\\
  g_0 &= \vec{u}^* \mdot \vec{f}_{1010}, &
  g_1 &= \vec{u}^* \mdot \vec{f}_{2100}, &
  g_2 &= \vec{u}^* \mdot \vec{f}_{1020}, &{}&{} \\
  c_0 &= \vec{v}^* \mdot \vec{f}_{1100}, &
  c_1 &= \vec{v}^* \mdot \vec{f}_{0020}, &
  c_2 &= \vec{v}^* \mdot \vec{f}_{1110}, &
  c_3 &= \vec{v}^* \mdot \vec{f}_{0030}.  \nonumber
\end{align}
The results derived for the fold-Hopf bifurcation are
summarized in \tab{tab:FoldHopfTable}.

The amplitude equation for a reaction-diffusion system that
undergoes a simple Hopf bifurcation is easily obtained from
the general result \eqn{eq:FinalAmp} following the same
procedure.  As a result, we obtain the well-known complex
Ginzburg-Landau equation (CGLE)
\begin{equation}
  \dot{w} = \sigma_1 \mu w + g \abs{w}^2w + d \nabn w.
  \label{eq:CGLE}
\end{equation}
The details of the results for the CGLE and explicit
expressions for calculations of the coefficients in
\eqn{eq:CGLE} are summarized in \tab{tab:HopfTable}.  The
parameter $d$ in~\eqn{eq:CGLE} is identical with $d_w$, shown
in~\eqn{eq:FoldHopf}.

\section{Handling a slow noncritical mode}
\label{sec:Handling}

An amplitude equation very similar to the distributed
fold-Hopf equation~\eno{eq:FoldHopf} can be derived for a
system with a Hopf bifurcation and a slow real mode, even if
only a Hopf bifurcation can be realized (near the desired
operating point or anywhere at all).  The result may be viewed
as an extension of the complex Ginzburg-Landau equation.

\subsection{The distributed slow-Hopf equation}
\label{sec:DSHE}
We consider a system depending on a scalar parameter $\mu$,
having a Hopf bifurcation (for the homogeneous system) at
$\mu=0$. The critical eigenvalues are $\pm \I\omega_0$ and the
corresponding eigenvectors are $\vua$ and $\vub$. All other
eigenvalues have negative real parts. One real eigenvalue is
small, $\abs{\lambda_0} \ll \omega_0$ (eigenvector $\vuc$)
whereas all others satisfy $-\re \{\lambda\} \gg \omega_0$.

We may describe the system with a partial unfolding of the
fold-Hopf equation, translated so that the origin is at the
stationary point at $\mu=0$.  Unfolding of the Hopf
bifurcation is governed explicitly by $\mu$ and needs no
further discussion.  The unfolding and translation from the
fold-Hopf bifurcation to the particular Hopf bifurcation
studied, can be accounted for by evaluating the fold-Hopf
parameters at the Hopf bifurcation point, $(\vec{x},\mu)
=(\vec{0},0)$ instead of at the (fictitious) fold-Hopf
point. In addition, the term $\rho _0\mu$ of \eqn{eq:FinalAmp}
disappears because of the translation to the new stationary
point, and a term $\lambda_0 z$ appears explicitly, which was
absent at the fold-Hopf bifurcation because $\lambda_0$
vanishes there. The relation of unfoldings to parameters of an
amplitude equation and amplitude transformation, evaluated at
a stationary point of an unfolded equation, has been discussed
in section VII of \cite{Ipsen98}.

The result of the modification is an equation of exactly the
same form as \eqn{eq:FinalAmp} except that $\lambda_0 z$
replaces $\rho_0\mu$. The coefficients are given by exactly
the same formulas, \tab{tab:FoldHopfTable}, except that the
Jacobian matrix and the other derivatives of the vector field
are evaluated at the Hopf bifurcation point, and so are the
eigenvalues and eigenvectors of $\vec{J}$. We shall refer to
the equation as a distributed slow-Hopf equation (DSHE).
Keeping the most important (lower order) terms only, the DSHE
takes the form
\begin{subequations}
  \label{eq:DSHE}%
  \begin{align} 
    \label{eq:DsheA}%
    \dot{w} &= \sigma_1 \mu w + g_0 w z + g_1 \abs{w}^2w + d_w \nabn w,\\
    \label{eq:DsheB}%
    \dot{z} &= \lambda_0 z + c_0 \abs{w}^2 + d_z \nabn z,
  \end{align}
\end{subequations}
which we use in \sect{sec:Comparison} to describe a realistic
model of the BZ reaction-diffusion system.

The DSHE \eno{eq:DSHE} is a generalization of the CGLE
\eno{eq:CGLE}.  It contains the simple CGLE as a special case
when $\abs{\lambda_0} \gg \re\{\sigma_1\}\mu$, as we show in
\app{sec:CGLERel}: there we discuss the relation between the
two equations, and in particular we derive an explicit
relation between the parameters $g_1$ and $g$.  These
parameters are in general quite different despite their
deceptively similar definitions. The two amplitude equations
are evaluated and compared numerically with each other and
with the underlying reaction-diffusion equation in
\sect{sec:Comparison}.

\subsection{Scalings}
\label{sec:Scalings}
Near a bifurcation, an evolution equation often scales with
the bifurcation parameter(s) to lowest order(s) in a simple
way.  This property admits a reduction of the (truncated )
equation to a form independent of the distance(s) from the
bifurcation point.  Examples of reduced equations are the
Ginzburg-Landau equation~\citep{Kuramoto}, the
Kuramoto-Sivashinsky equation~\cite{KT76,Si77}, and the
Swift-Hohenberg equation~\citep{CrossHohen}.

For the fold-Hopf bifurcation, a complete scaling requires two
carefully chosen bifurcation parameters.  The
DSHE~\eno{eq:DSHE} cannot be scaled to a form independent of
the bifurcation parameter.  Nevertheless, it is useful to
introduce a partial rescaling of \eqn{eq:DSHE} with the
substitutions
\begin{equation}
      w  = \sqrt{\mu} w',  \qquad
      z  = \mu  z',        \qquad
      t  = t'/\mu,         \qquad
 \vec{r} = \vec{r}'/\sqrt{\mu} 
  \label{eq:CGLEScaling}
\end{equation}
to obtain an equation in the new, primed variables.  If we
drop the primes for simplicity, the scaled equation becomes
\begin{subequations}
  \begin{align}
    \dot{w} &= \sigma_1 w + g_0 w z + g_1 \abs{w}^2w + d_w \nabn w,\\
 \mu\dot{z} &= \lambda_0 z + c_0 \abs{w}^2 + \mu d_z \nabn z.
  \end{align}
  \label{eq:FoldHopfScaled}%
\end{subequations}

For analysis of solutions and numerical computations it is
convenient to reduce \eqn{eq:FoldHopfScaled} to dimensionless
form by further scalings through the substitutions
\begin{equation}
        w  = \sqrt{\frac{-\sgr}{\grb}} w'
           \exp (\I\frac{\sgi}{\sgr}t'), \qquad
        z  = \frac{\sgr}{\gra} z', \qquad
        t  = t'/\sgr, \qquad
  \vec{r}  = \sqrt{\frac{\drb}{\sgr}} \vec{r}',
\end{equation}
in which superscripts $\mathrm{r}$ and $\mathrm{i}$ denote
real and imaginary parts respectively.  The result is an
equation in the primed variables.  If we skip the primes for
convenience, the resulting rescaled equation takes the form
\begin{subequations}
  \begin{align}
    \dot{w} &= w + (1 + \I\gamma)wz - (1 + \I\ash) w\abs{w}^2 + 
              (1 + \I\beta) \nabn w,\\
    \epsilon
    \dot{z} &= \lambda_0 z + \kappa \abs{w}^2 + \epsilon\delta \nabn z,
  \end{align}
  \label{eq:ScaledFold}%
\end{subequations}
where 
\begin{equation}
      \ash = \frac{\gib}{\grb},      \qquad
    \beta  = \frac{\dib}{\drb},      \qquad
    \gamma = \frac{\gia}{\gra},      \qquad
  \epsilon = \mu\sgr,                \qquad
    \kappa = -c_0 \frac{\gra}{\grb}, \qquad
    \delta = \frac{d_z}{\drb}.
    \label{eq:FHPars}
\end{equation}
The scalings used here are similar to those traditionally used
for the CGLE \citep{Kuramoto} reducing it to the form
\begin{equation}
    \dot{w} = w - (1 + \I\alpha) w\abs{w}^2 + (1 + \I\beta) \nabn w.
  \label{eq:DimCGLE}
\end{equation}
Here $\beta$ is given by \eqn{eq:FHPars} and $\alpha =
\tfrac{g^\mathrm{i}}{g^\mathrm{r}}$.  The parameters $\alpha$
and $\ash$ differ because $g$ and $g_1$ do, as discussed
further in \app{sec:CGLERel}.

\section{Comparison of solutions}
\label{sec:Comparison}
The first successful attempt to model the oscillatory behavior
of the BZ reaction was the Oregonator model suggested by Field
and Noyes~\citep{OregOrig}.  It exhibits both sinusoidal and
relaxational oscillations in semi-quantitative agreement with
many experiments.  However, the Oregonator model does
\emph{not} reproduce more complex dynamics such as
quasiperiodic oscillations or chaos.

\subsection{Model of the BZ reaction}
\label{sec:BZModel}
To incorporate complex phenomena of the BZ reaction into the
modeling many different variants of the Oregonator have been
considered.  Here we use the modified Oregonator recently
suggested by Wang et.\ al~\cite{Wang95} to explain very
complicated transient phenomena observed in the BZ reaction.
The model, which we refer to as the 4D~Oregonator, is based
upon the following chemical scheme
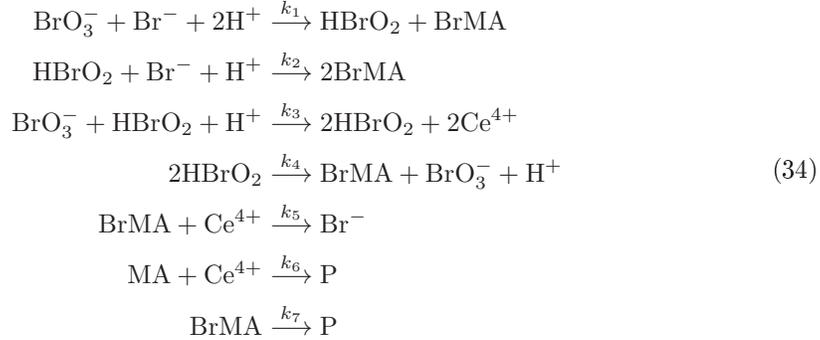
\begin{equation}
  \begin{split}
    \bromat + \bromid + 2\hp & \stackrel{k_1}{\lra} \brsyr + \brma\\
    \brsyr  + \bromid + \hp  & \stackrel{k_2}{\lra} 2\brma\\
    \bromat + \brsyr + \hp   & \stackrel{k_3}{\lra} 2\brsyr + 2\cef\\ 
    2\brsyr                  & \stackrel{k_4}{\lra} \brma + \bromat + \hp\\
    \brma + \cef             & \stackrel{k_5}{\lra} \bromid\\
    \ma + \cef               & \stackrel{k_6}{\lra} \chem{P}\\
    \brma                    & \stackrel{k_7}{\lra} \chem{P}
  \end{split}
  \label{eq:4DOrgReac}
\end{equation}
Note that the first six reactions are equivalent with the
reactions in the Oregonator model, except for the fact that
bromomalonic acid (\brma) appears explicitly in the scheme.
The mechanistic assumption is that \chem{HBrO} reacts
immediately with malonic acid to produce \brma.  In order to
explain the complexity observed in the closed system it is
essential to add a reaction which corresponds to a removal of
\chem{\brma} without a simultaneous production of
\chem{\bromid}.  This additional feature is included in
reaction 7 of \eqn{eq:4DOrgReac}.

Now, if we regard the concentration of \brma~as an additional
dynamical variable and assume non-homogeneous spatial
conditions, the above set of reactions gives rise to the
following four-dimensional differential equation
\begin{equation}
  \begin{split}
    \frac{\partial X}{\partial t} &=
     k_1 A H^2 Y - k_2 H X Y + k_3 A H X - 2 k_4 X^2 +
    D_X\nabla_\vec{r}^2X, \\[2.0mm]
    \frac{\partial Y}{\partial t} &=
    -k_1 A H^2 Y - k_2 H X Y + k_5 U Z +
    D_Y\nabla_\vec{r}^2Y, \\[2.0mm]
    \frac{\partial Z}{\partial t} &=
    2 k_3 A H X - k_5 U Z - k_6 M Z +
    D_Z\nabla_\vec{r}^2Z, \\[2.0mm]
    \frac{\partial U}{\partial t} &=
    k_1 A H^2 Y + 2 k_2 H X Y + k_4 X^2 - k_5 U Z - k_7 U +
    D_U\nabla_\vec{r}^2U,
  \end{split}
  \label{eq:Diff4DOrg}
\end{equation}
where the following short-hand notation is introduced for the
concentration of the chemical species: $X = \chem{[\brsyr}]$,
$Y = \chem{[\brom]}$, $Z = \chem{[\cef]}$, $U =
\chem{[\brma]}$, $A = \chem{[\bromat]}$, $H = \chem{[\hp]}$,
and $M = \chem{[MA]}$.  As in the Oregonator model, we regard
$A$ as a constant in order to study the system under open
conditions.  Besides having a supercritical Hopf bifurcation
with small sinusoidal oscillations, the model also contains an
additional slow time scale which primarily is related to the
kinetics of \brma.  The values of the various constants in
\eqn{eq:Diff4DOrg} are shown in \tab{tab:Diff4DOrg}.

\subsection{Bifurcations and stability of waves}
\label{sec:Bif-Waves}
For the comparisons, we need to select a suitable operating
point for the 4D~Oregonator, close to a supercritical Hopf
bifurcation.  As parameters, we have chosen the fixed
concentrations \chem{[\bromat]} and \chem{[\hp]}.  All other
parameters have been fixed using the values shown
in~\tab{tab:Diff4DOrg}.

The locus of Hopf bifurcations in the plane of the free
parameters are shown in \figa{fig:Hopf4DDiff}{a}. The
bifurcation diagram consists of two apparently separate
curves, which are connected for a very large and physically
unrealistic value of \chem{[\bromat]}, however. Both branches
describe supercritical Hopf bifurcations. The filled circle on
the upper branch indicates a Hopf bifurcation point which has
been chosen as reference point \emph{viz.}
\chem{[\bromat]_{Hopf} = 0.3662\,\molar} and
\chem{[\hp]_{Hopf} = 1.3416\,\molar}.  This reference point is
characterized in~\tab{tab:HopfData}.  Since it is necessary to
work at a non-zero distance from the bifurcation point, the
actual operating point will be chosen close to this reference
point.  The variation of the Ginzburg-Landau parameters
$\alpha$ and $\beta$, and the functional dependence
$\beta(\alpha)$ are shown in
Figs.~\ref{fig:Hopf4DDiff}b--\ref{fig:Hopf4DDiff}d. In all
three cases, numbers \ding{192}--\ding{194} identify curves
with the branches in the bifurcation diagram.

The reference point shown with a filled circle in
\fig{fig:Hopf4DDiff} has been chosen with a view to the
stability properties of waves.  Reaction-diffusion equations
may support several types of waves.  Near a supercritical Hopf
bifurcation, in particular, there may exist stable spiral
waves as well as spatio-temporal chaos (chemical turbulence),
depending on parameter conditions.  Usually the CGLE
approximation faithfully reproduce such
behavior~\citep{Ipsen97}, and it is convenient to relate wave
stability to the parameters, $\alpha$ and $\beta$, of the
CGLE.

For the CGLE, the stability of spiral waves and their
transitions to turbulent solutions in the
$\alpha,\beta$-parameter plane have been studied numerically
and theoretically in \citep{Huber92,Ar92a,Ar92b}.  One key
observation in these studies is that the transition to
turbulence in the CGLE is closely related to the concepts of
convective and absolute instability.  For spiral wave
solutions, it was found that the boundary of the region for
absolute instability is very close to the curve where spiral
wave solutions exhibits a transition to turbulence.  In
addition, the region which separates the onset of convective
instability of the spiral waves, to some extent, may be
approximated by the Eckhaus instability border
\citep{Ar92a,Ar92b}.  The parameter space may therefore be
divided into several subregions described by convective and
absolute instabilities as in \figa{fig:Hopf4DDiff}{d}.  In the
next section, we shall use these concepts in a discussion of
suitable operating points to be selected for a comparison of
the various equations, set up for the 4D Oregonator.


\subsection{CGLE versus the reaction-diffusion equation}
\label{sec:CGLEDes}
We choose the operating point well outside the region of
absolute instability (AI in \figa{fig:Hopf4DDiff}{a}) where we
expect to observe stable spiral waves for the CGLE.  Observe,
that the chosen point lies close to the region of convective
instability (EH), but this has no importance to the local
stability properties of spiral waves.  The Ginzburg-Landau
parameters evaluated at the reference point are shown in
\tab{tab:CGLEData}.

The 4D~Oregonator must be run at a non-zero distance from the
Hopf bifurcation point.  We choose the parameter values
\chem{[\bromat] = 0.3625\,\molar} and \chem{[\hp] =
[\hp]_{Hopf}}. This corresponds to an amplitude of the uniform
oscillations, which is 7.5\% of the stationary \chem{\cef}
concentration.  We emphasize that the CGLE is scaled and hence
independent of the distance from the bifurcation point but the
reverse scaling together with the amplitude transformation
account for the distance actually used.

The most systematic way of comparing the solutions to the
amplitude equation and the reaction-diffusion equation is to
transform the space and time dependent complex amplitudes back
to concentration space using the scalings described in
\sect{sec:Scalings} and the amplitude transformation
determined in \tab{tab:HopfTable}.  The transformation is
nonlinear, and the nonlinear terms are sometimes important
\cite{MiIs98}. 

For the present study, the linear part of the transformation 
\begin{equation}
  \vec{x}(\vec{r},t;\mu) = \statb{x}(\mu) + 
  \Bigl[
    w(\vec{r},t;\mu) \E^{\I\omega_0 t} \vua + \mathrm{c.c.}
  \Bigr], 
  \label{eq:Obvious}
\end{equation}
is sufficiently accurate. \Eqn{eq:Obvious} is expressed in
terms of unscaled variables $w$, $\vec{r}$, and $t$.  In
addition, the unfolding terms describing the variation of the
stationary point $\statb{x}(\mu)$ with the parameter $\mu$ are
also incorporated.  As an alternative to transforming
amplitudes to physical variables for the comparison, one may
calculate the amplitudes from the physical variables
(concentrations) obtained from the reaction-diffusion
equation.  This is easy in the linear approximation to the
amplitude transformation:
\begin{equation}
  w(\vec{r},t;\mu) = 
  \vua^* \mdot  
  \Bigl[
    \vec{x}(\vec{r},t;\mu) - \statb{x}(\mu) 
  \Bigr] \E^{-\I\omega_0 t}.
  \label{eq:NotObvious}
\end{equation}
The representation \eno{eq:NotObvious} is very convenient
since the magnitude $\abs{w}$ of the (unscaled) amplitude
forms a very clear representation of an oscillatory wave.

In the comparisons, \figs{fig:ReacAndGinzA}{fig:ReacAndGinzB},
we exhibit the magnitude $\abs{w(\vec{r},t;\mu)}$ as well as
selected components of $\vec{x}(\vec{r},t;\mu)$ (the
concentrations of \cef and \brma) for solutions to the
reaction-diffusion equation (RDE) and the CGLE respectively,
using initial conditions suggested by Kuramoto~\cite{Kuramoto}
(we choose $\abs{w(\vec{r},t;\mu)}$ proportional to
$\abs{\vec{r}}$ and $\arg(w(\vec{r},t;\mu)) = \arg(\vec{r})$,
the polar angle).


As is clear from \fig{fig:ReacAndGinzB}, the CGLE exhibits the
behavior expected from the diagram \figa{fig:Hopf4DDiff}{d}, a
spiral wave solution evolving from the center of the grid.
However, this very regular solution shows virtually no
resemblance with that of the reaction-diffusion equation,
\fig{fig:ReacAndGinzA}.  Focusing on $[\cef]$, we see that a
small spiral is initially formed at the center of the
grid. After a few windings, the spiral is no longer able to
maintain its structure, and a turbulent pattern develops.
This is especially clear in the amplitude plot where a ``shock
wave'' is emitted from the spiral core. When the shock hits
the boundary of the domain, many small spirals are generated,
the symmetry of the initial state breaks, and fully developed
turbulence is reached.  Here numerous small spirals are
constantly created and annihilated, resulting in a disordered
motion typical of chemical turbulence, which also has been
observed in certain parameter regions for a reaction-diffusion
equation based on the simple three dimensional Oregonator
model~\citep{Ipsen97}.



\subsection{DSHE versus the reaction-diffusion equation}
\label{sec:DSHEDes}
The failure of the CGLE is clearly a result of the presence of
a slow real mode in the 4D~Oregonator, which violates a
condition underlying a derivation of the equation: all motion
towards the plane of the oscillations must be fast compared
with the motion in the plane.  We shall therefore try to
describe the reaction-diffusion system with a distributed
slow-Hopf equation, DSHE, which takes the slow real mode
explicitly into account.

We work at the same operating point as used in the previous
section so the results can be immediately compared with those
for the CGLE and the reaction-diffusion equation.  The
parameters of the DSHE are evaluated at the Hopf bifurcation
point using the formulas of \tab{tab:FoldHopfTable}.  They are
shown in \tab{tab:DSHEData}.  Note that the parameter $g_1$ is
quite different from the corresponding CGLE coefficient $g$.
This important difference is discussed in \app{sec:CGLERel}.

We want to compare the solution to the DSHE with that of the
reaction-diffusion equation first of all.  For the latter, we
might use \fig{fig:ReacAndGinzA}, but to illuminate the role
of the slow real mode, we shall calculate the real amplitude
$z$ as well as the concentrations of $\cef$ and \brma.

The concentrations are obtained from the scaled amplitudes by
reverse scaling and transformation back to concentration space
with the amplitude transformation described in
\tab{tab:FoldHopfTable}.  As before, we consider only the
linear part of the transformation here:
\begin{equation}
  \vec{x}(\vec{r},t;\mu) = \statb{x}(\mu) + 
  \bigl(
    w(\vec{r},t;\mu) \E^{\I\omega_0 t} \vua + \mathrm{c.c.}
  \bigr) +
  z(\vec{r},t;\mu) \vuc,
  \label{eq:ExObvious}
\end{equation}
(shown in terms of unscaled variables).  In the linear
approximation \eno{eq:ExObvious}, we may alternatively
calculate the amplitudes from a solution to the
reaction-diffusion equation as
\begin{subequations}
  \begin{align}
    \label{eq:ExNotA}%
    w(\vec{r},t) &= 
    \vua^* \mdot  
    \Bigl[
    \vec{x}(\vec{r},t;\mu) - \statb{x}(\mu) 
    \Bigr] \E^{-\I\omega_0 t},\\
    \label{eq:ExNotB}%
    z(\vec{r},t) &= 
    \vuc^* \mdot  
    \Bigl[
    \vec{x}(\vec{r},t;\mu) - \statb{x}(\mu) 
    \Bigr].
  \end{align}\\
  \label{eq:ExNotObvious}%
\end{subequations}
The linear approximation \eno{eq:ExNotA} to the complex
amplitude $w$ is identical to the one calculated for the CGLE,
\eqn{eq:NotObvious} (but the two amplitudes differ if
nonlinear terms of the amplitude transformation are taken into
account).  Its magnitude $\abs{w}$ has already been shown in
\fig{fig:ReacAndGinzA}, but to save space, we shall not
exhibit $\abs{w}$ for the DSHE\@.  Instead, we show the real
amplitude z together with $[\cef]$ and [\brma] in
\fig{fig:Ex4DCompA} for the reaction-diffusion equation and
\fig{fig:Ex4DCompB} for the DSHE\@.  Initial conditions for
the simulations are the same as in the previous section
\cite{Kuramoto}.

The agreement between the two solutions is striking.  Even
quite subtle details agree for a long time although,
eventually, the details of the turbulent patterns must
disagree.  However, the general character of the solutions
continue to agree.  Thus, the improvement of the
DSHE-approximation over that of the CGLE,
\fig{fig:ReacAndGinzB}, is evident.

In conclusion, it is possible to treat a reaction-diffusion
problem near a Hopf bifurcation using an efficient amplitude
description, even for systems like the BZ-reaction that
contain a very slow real mode as an intrinsic part of the
chemistry.  Essentially, the DSHE treats the Hopf bifurcation
(of codimension one) as an unfolded fold-Hopf bifurcation of
codimension two.

\subsection{Role of the slow real mode}
\label{sec:SlowMode}
A conspicuous feature of the solutions to the
reaction-diffusion equation, \fig{fig:Ex4DCompA}, or the DSHE,
\fig{fig:Ex4DCompB}, is the similarity of the patterns for
[\brma] and the amplitude $z$.  In fact, the two patterns are
almost identical either in \fig{fig:Ex4DCompA} or in
\fig{fig:Ex4DCompB}.

This similarity can be explained in terms of the eigenvectors
of the Jacobian matrix whose numerical values, evaluated at
the stationary point at the Hopf bifurcation, are shown in
\tab{tab:EigData}.  We note that the eigenvector $\vec{v}$ for
the slow real mode is directed almost exactly along the
$[\brma]$-axis.  

Another noticeable feature of the solutions in
\figs{fig:Ex4DCompA}{fig:Ex4DCompB} is the different
characters of the components $[\brma]$ and $[\cef]$.  Clearly,
the spatio-temporal evolution of the \brma~pattern resembles
that of $\abs{w}$ much more (see~\fig{fig:ReacAndGinzA}).
Again, the eigenvectors, \tab{tab:EigData}, can throw some
light on this feature.  In the linear approximation, the left
eigenvectors $\vua^*$ and $\vub^*$ (or the real and imaginary
parts of $\vua^*$) define the projection of any vector onto
the plane of oscillation.  From \tab{tab:EigData} we see that
the $[\brma]$ ``component'' of $\vua^*$ is very small compared
with the other components of $\vua^*$.  This implies that
$[\brma]$ virtually does not participate in the basic
oscillations of frequency $\omega_0$.  Thus, apart from scale,
the amplitude $z$ can almost be identified with
$\chem{[BrMA]-[BrMA]_0}$.

To understand the behavior of $[\brma]$ and the role of the
amplitude $z$, we first watch the local variations of $\re w$
and $z$ at an interior grid point from the simulation shown in
\fig{fig:Ex4DCompB}.  We observe that the amplitude $R$ of the
oscillations $\re w$ appear to vary in roughly the same way as
$z$.

Setting $w = R \exp(\I\Theta)$, gives the following
differential equation for $R$ and $z$ in the homogeneous DSHE
\begin{subequations}
  \label{eq:DsheAmp}
  \begin{align}
    \dot{R} &= (\sgr + \rin{g_0}{r}z)R - \rin{g_1}{r}R^3,\\
    \mu\dot{z} &= \lambda_0 z + c_0 R^2,
  \end{align}
\end{subequations}
where superscript $\mathrm{r}$ denotes the real part.  To put
\eqn{eq:DsheAmp} into dimensionless form we introduce the
variable change
\begin{equation}
  R = \sqrt{-\frac{\sgr}{\rin{g_1}{r}}} R', \qquad 
  z = \frac{\sgr}{\rin{g_0}{r}} z', \qquad \text{and} \qquad
  t = \frac{1}{\sgr} t'.
\end{equation}
Skipping the primes for simplicity then gives the following
dimensionless equation
\begin{subequations}
  \label{eq:DsheAmp2}
  \begin{align}
    \dot{R} &= R + Rz - R^3,\\
    \dot{z} &= \frac{\lambda_0 z + \kappa R^2}{\mu\sgr},
    \quad \text{where} \quad \kappa = -c_0 \frac{\rin{g_0}{r}}{\rin{g_1}{r}}.
  \end{align}
\end{subequations}

Apart from $(0,0)$, this scaled differential equation admits
the non-trivial stationary solution
\begin{equation}
    \stat{R} = \bigl( 1 + \frac{\kappa}{\lambda_0} \bigr)^{-\frac{1}{2}},\quad
    \stat{z} = -\frac{\kappa}{\kappa + \lambda_0},
\end{equation}
which corresponds to a uniform oscillatory solution to the
reaction-diffusion equation. A numerical solution to
\eqn{eq:DsheAmp2} is shown in \fig{fig:NullCl}.  Here, we
observe that the point $(\stat{R},\stat{z})$ corresponds to a
stable focus in the $(R,z)$-plane---the trajectory exhibits a
spiraling motion converging to the stationary state in
agreement with the oscillatory modulation of the amplitude
observed in \fig{fig:SteadyTrans}.  If we include diffusion
into the description, this will serve as a constant
perturbation, which keeps the solution away from the
stationary solution $(\stat{R},\stat{z})$ and thereby
preserves the modulation of the amplitude. The interaction
between the amplitudes $R$ and $z$ therefore introduces an
additional frequency into the system, as observed in the
torus-like behavior of the time series for $\re w$, and
destroys the stability of spirals.

\section{Conclusion}
\label{sec:Conclusion}
We have derived an amplitude equation that provides an
efficient and accurate, approximate description of a
reaction-diffusion system near a fold-Hopf bifurcation, as
well as generalizations to systems with several critical real
modes.  The derivation is based on a normal form
transformation combined with the use of two distinct time
variables to handle slow modulations of oscillations.  It does
not rely on different time scales.

The derivation exhibits a simple geometrical interpretation of
the amplitude equation in terms of the center manifold and
center subspace, and it provides an explicit relation, the
amplitude transformation, between amplitudes and physical
variables.  An amplitude equation is obtained as a truncation
of a form containing infinite series, and there is a choice of
including more or less terms.  Similarly one may include more
or less terms in the amplitude transformation.  Observe that
all coefficients can be calculated from the results of the
theory.  

The main result of the paper is the general amplitude equation
\eqn{eq:FinalAmp} for a Hopf bifurcation and any number of
``semisimple'' critical real modes.  It contains all the
higher order terms mentioned, and in practice, it must be
suitably truncated for any particular use.  We note in
particular that \eqn{eq:FinalAmp} includes nonlinear diffusion
terms.  

The special case of a fold-Hopf bifurcation is treated more
explicitly in \sect{sec:FoldHopfDer}.  Here the amplitude
equation~\eno{eq:FoldHopf} is a selected truncation of
\eqn{eq:FinalAmp} for the case of a single real critical mode.
The coefficients~\eno{eq:FoldHopfCoeff} are given explicitly
by \tab{tab:FoldHopfTable} in terms of coefficients
$\vec{h}_\vec{pq}$ obtained by solving the linear equations
shown.

The fold-Hopf bifurcation has codimension two, and we would
not expect to find a realization of this bifurcation in any
particular reaction-diffusion system.  Nevertheless, we show
that a description based on the dynamics of a fold-Hopf
bifurcation is much more versatile than the codimension
initially suggests.

We show that the machinery used to derive the distributed
fold-Hopf equation~\eno{eq:FoldHopf} can be used to derive
what we have called the distributed slow-Hopf
equation~\eno{eq:DSHE}, the DSHE, applicable to a
reaction-diffusion system with a simple Hopf bifurcation
together with a slow (near-critical) real mode.  The
DSHE~\eno{eq:DSHE} differs from the general amplitude
equation~\eno{eq:FinalAmp} by including the linear term
$\lambda_0 z$ which is absent at criticality.  In addition,
all its coefficients are evaluated at the Hopf bifurcation.
It includes terms sufficient for the particular
reaction-diffusion problem discussed: a four-dimensional model
of the Belousov-Zhabotinsky (BZ) reaction exhibited in
\sect{sec:BZModel}, the 4D Oregonator.


The purpose of looking at the 4D~Oregonator~\eno{eq:Diff4DOrg}
is first of all to test the DSHE by comparing solutions to it
with solutions to the reaction-diffusion equation and to the
corresponding CGLE, all computed at the same operating point
of~\eqn{eq:Diff4DOrg}.



The conclusion from the comparison
(\figs{fig:Ex4DCompA}{fig:Ex4DCompB}) is that the DSHE
describes the reaction-diffusion problem almost quantitatively
in the present case.  In contrast, the CGLE fails completely
(\figs{fig:ReacAndGinzA}{fig:ReacAndGinzB}).  So using the
DSHE, one may reduce the numerical work to a small fraction
(typically of the order 1:100) of that required for the
corresponding direct integration of the reaction-diffusion
equation.  For the operating point considered, the behavior of
the CGLE agrees with the fact that the coefficients $\alpha$
and $\beta$ of the equation lie in a region where plane wave
solutions are locally stable to long wave perturbations.  This
spurious stability is a consequence of the neglect of all
dynamics out of the plane of oscillations.  The correct
behavior, as represented by the DSHE, can be explained by a
stability analysis of plane wave solutions to the
DSHE~\eno{eq:DSHE} which we have carried out using a
modification of the approach described in \cite{Ar92a}.  We
shall report these results in a separate paper.

Another reason for treating the particular reaction-diffusion
system~\eno{eq:Diff4DOrg}, in terms of the DSHE,
\eqn{eq:DSHE}, is the importance of the BZ-reaction, which it
models.  The BZ-reaction is perhaps the most studied nonlinear
chemical reaction, and happens to have a very slow
near-critical real mode almost everywhere in parameter space.
(In fact, we have not found a fold-Hopf bifurcation anywhere
in the model.)  Furthermore, the slow real mode is very
important to the dynamics, being associated with complex
oscillations and chaos.  So for a realistic treatment of a
BZ-based reaction-diffusion system, the slow real mode cannot
be neglected.

Recently, qualitative comparisons between solutions to the
CGLE and experimentally observed spiral waves in the ferroin
catalyzed BZ reaction were reported by
Ouyang~and~Flesselles~\cite{CgleNature}.  The ferroin
catalyzed BZ reaction is not so well understood and
characterized as the \cef catalyzed version.  Although a large
part of the reactions involving bromomalonic acid are common
for the two versions of the BZ system, reactions like the
fifth of \eqn{eq:4DOrgReac} do depend on the catalyst.
Nevertheless, the possibility of slow real modes should be
kept in mind, and the applicability of the CGLE cannot be
taken for granted.

An advantage of the CGLE is that its coefficients $\alpha$ and
$\beta$ can be obtained directly from
experiments~\cite{PgsCgle} although the methods still need
some refinement for higher accuracy.  Unfortunately, measuring
the coefficients of \eqn{eq:DSHE} is much more difficult.  We
may possibly get some progress using more general perturbation
methods based on control theory, which we presently are
developing.  Otherwise one must work through optimized models
of the reaction considered.

In a case where several real modes are close to criticality,
one must derive the appropriate amplitude equation from the
general expression~\eno{eq:FinalAmp}.  This means calculating
explicit expressions for the necessary coefficients as in
\tab{tab:FoldHopfTable}.  At the same time the linear
equations for the coefficients of the amplitude transformation
must be solved to the order desired.  These steps are quite
straightforward using software packages capable of performing
symbolic manipulations such as Mathematica or
Maple~\cite{SW96,Maple}.  Even though the multitude of
coefficients to be calculated and equations to be solved grows
rapidly with the number of slow modes, the gain in speed by
using an amplitude equations may well make a solution feasible
that in practice is impossible using the reaction-diffusion
equation directly.

\appendix

\section{Amplitude transformation}
\label{app:AmpTrans}
In \cite{Ipsen98}, we derived an expression for the unfolded
center manifold $\Wc(\bmu)$ in terms of coordinates $y_i$ of
$\vec{y} \in \Ec$ in a basis of critical eigenvectors
$\vec{u}_i$, namely
\begin{equation}
  \vec{x} = 
  \vec{y} + \vec{h}(\vec{y},\bmu) =
  \sum_{i=1}^r y_i \vec{u}_i +
  \sum_\vec{pq} \vec{h}_\vec{pq} \vec{y}^\vec{p}\bmu^\vec{q}
  \label{eq:AppCenterExp}
\end{equation}
in which $\vec{y}^\vec{p} = \prod_i y_i^{p_i}$ and
$\bmu^\vec{q} = \prod_k \mu_k^{p_k}$.  The coefficient vectors
$\vec{h}_\vec{pq}$ are solutions to the linear equations
\begin{subequations}
  \label{eq:LinearEqUn}
  \begin{align}
    \label{eq:LinearEqUnA}
    \Bigl(
      \vec{J}  - \sum_{j=1}^r p_j \lambda_j \vec{I}
    \Bigr) \mdot 
    \vec{h}_\vec{pq} &= 
    -\vec{Q}_\vec{p} \mdot \BPhi_\vec{pq},\\
    \vec{R}_\vec{p} \mdot \vec{h}_\vec{pq} &= \vec{0}.
    \label{eq:LinearEqUnB}
  \end{align}
\end{subequations}
in which $\vec{I}$ is the unit tensor, $\vec{R}_\vec{p}$
projects onto the resonant subspace and $\vec{Q}_\vec{p}$
removes the resonant components of any vector~$\vec{x}$:
\begin{equation}
  \label{eq:ProjDef}
  \vec{R}_\vec{p} \mdot \vec{x} = 
  (\vec{I} - \vec{Q}_\vec{p}) \mdot \vec{x} =
  \sum_i
  (\vec{u}_i^* \mdot \vec{x}) \vec{u}_i.
\end{equation}
The sum in \eqn{eq:ProjDef} is taken over all components $i$
for which the resonance condition~\eno{eq:ResDiff} is
satisfied (for a given order $\vec{p}$).  The ``source term''
$\BPhi_\vec{pq}$ on the right-hand side of \eqn{eq:ProjDef} is
given by the expression
\begin{equation}
  \BPhi_\vec{pq} = \vec{f}_\vec{pq} -
  \sum_{\vec{p}'\vec{q}'} 
  \vec{h}_{\vec{p}'\vec{q}'} 
  \sum_{j=1}^r p_j'
  \vec{u}_j^* \mdot
  \vec{g}_{(\vec{p}-\vec{p}'+\bdl_{\!j})(\vec{q}-\vec{q}')}.
  \label{eq:PhiDefUnfold}
\end{equation}
in terms of $\vec{f}_{\vec{pq}}$ and lower order coefficients
$\vec{h}_{\vec{p}'\vec{q}'}$ and
$\vec{g}_{\vec{p}''\vec{q}''}$, given by
\begin{equation}
  \vec{u}_j^* \mdot 
  \vec{g}_{\vec{p}''\vec{q}''} = 
  \begin{cases}
    \vec{u}^*_j \mdot \vec{f}_{\vec{p}''\vec{q}''}
    & \text{if $\vec{p}'' \mdot \bla = \lambda_j$},\\
  0 & \text{otherwise}.
  \end{cases}
  \label{eq:Udot}
\end{equation}

The vectors $\vec{f}_\vec{pq}$ are coefficients in an expansion of
the vector field $\vec{f}(\vec{x},\bmu)$ on the unfolded center
manifold $\Wc(\bmu)$,
\begin{equation}
  \vec{f}(\vec{y} + \vec{h}(\vec{y},\bmu),\bmu) =
  \sum_\vec{pq} \vec{f}_\vec{pq} \vec{y}^\vec{p}\bmu^\vec{q}.
  \label{eq:FExp}
\end{equation}
In practice, this expansion is obtained by substituting the
expansion~\eno{eq:AppCenterExp} in a Taylor expansion of
$\vec{f}(\vec{x},\bmu)$ in terms of the usual multilinear
forms, \emph{e.g.}
\begin{equation}
  \begin{split}
    \Dfxx(\vec{x},\vec{x}) &= 
    \sum_{i,j=1}^{n} 
    \frac{\partial^2 \vec{F}}{\partial x_i\partial x_j} x_ix_j,\qquad
    \Dfxxx(\vec{x},\vec{x},\vec{x}) = 
    \sum_{i,j,k=1}^{n} 
    \frac{\partial^3 \vec{F}}{\partial x_i\partial x_j\partial x_k} 
    x_ix_jx_k,\\
    \vec{F}_{\bmu} \mdot \bmu 
    &= \sum_{i=1}^{s} \frac{\partial \vec{F}}{\partial \mu_i}\mu_i,\qquad
    \vec{F}_{\vec{x}\bmu}(\vec{x},\bmu) = 
    \sum_{i=1}^{n} \sum_{j=1}^{s} \frac{\partial^2 \vec{F}}
    {\partial x_i\partial \mu_j} x_i\mu_j, \quad \text{etc.}
  \end{split}
  \label{eq:CompExp}
\end{equation}

Any coefficient $\vec{f}_\vec{pq}$ of the
expansion~\eno{eq:FExp} is then obtained as an expression that
may involve coefficients $\vec{h}_{\vec{p}'\vec{q}'}$.  The
important point now is that only lower orders
$(\vec{p}',\vec{q}')$ appear for $\vec{h}_{\vec{p}'\vec{q}'}$
in $\vec{f}_\vec{pq}$.  So it is always possible to
solve~\eqnto{eq:LinearEqUn}{eq:PhiDefUnfold} order by order.

As a result, the amplitude
transformation~\eno{eq:AppCenterExp} is obtained explicitly.
This procedure provides a particular representation of the
unfolded center manifold $\Wc(\bmu)$ in terms of the
amplitudes $y_i$.  Some simple examples of the calculation of
the coefficients $\vec{f}_\vec{pq}$ and explicit solutions of
the linear equation~\eno{eq:LinearEqUn} can be found in
\cite{Ipsen98}.

\section{Proof of \eqns{eq:Average}{eq:ParZero}}
\label{app:AverageZero}
We use the expansion in \eqn{eq:AppCenterExp} for
$\vec{h}(\vec{y},\bmu)$ together with \eqn{eq:YProd}.  For the
left-hand side of \eqn{eq:Average}, we then get
\begin{equation}
  \sum_\vec{pq} \vec{u}_i^* \mdot \vec{h}_\vec{pq} 
  \vec{z}^\vec{p} \bmu^\vec{q}
  \sum_{j=1}^r \frac{p_j}{z_j} \diff{z_j}{\tau}
  \frac{1}{T} \int_0^T \Exi d\theta.
\end{equation}
Here a term of order $(\vec{p},\vec{q})$ vanishes because of
the integral over $\theta$ unless $\vec{p} \mdot \bla =
\lambda_i$.  So the sum contains resonant terms only, but each
of these vanishes because resonant components of
$\vec{h}_\vec{pq}$ vanish, $\vec{u}_i^* \mdot \vec{h}_\vec{pq}
= 0$, as \eqn{eq:AppCenterExp} and \eqn{eq:LinearEqUn} show.
This demonstrates the identity in \eqn{eq:Average}.

For \eqn{eq:ParZero}, we use \eqn{eq:AppCenterExp} and first
observe that
\begin{equation}
  \vec{J} \mdot \vec{y} - \diff{\vec{y}}{\theta} =
  \vec{J} \mdot \Exp \mdot \vec{z} - 
  \diff{}{\theta}(\Exp \mdot \vec{z}) = 0
\end{equation}
From the expansion~\eno{eq:AppCenterExp} of the transformation
$\vec{h}(\vec{y},\bmu)$, we therefore get from the left-hand
side of \eqn{eq:ParZero}
\begin{equation}
  \sum_\vec{pq} \vec{u}_i^* \mdot\, (\vec{J} - \vec{p} \mdot \bla)
  \mdot \vec{h}_\vec{pq}
  \vec{z}^\vec{p} \bmu^\vec{q}
  \frac{1}{T} \int_0^T \Exi d\theta.
\end{equation}
Here only resonant terms contribute to the sum because of the
integral, but these also vanish since $\vec{u}_i^* \mdot
(\vec{J} - \vec{p} \mdot \bla) = 0$.  The whole sum is
therefore identically zero, so \eqn{eq:ParZero} is proved.

\section{Relation to the CGLE}
\label{sec:CGLERel}
Comparsion of the DSHE~\eno{eq:DSHE} with the
CGLE~\eqn{eq:CGLE} contains a subtlety, which we now discuss.
\Eqn{eq:DsheA} has the same form as \eqn{eq:CGLE} apart from
the coupling term $g_1 w z$.  But the coefficients $g_1$ and
$g$ also differ.  They are defined in
\tabs{tab:FoldHopfTable}{tab:HopfTable} by very similar
expressions, namely
\begin{align}
  \label{eq:GExpA}
  g_1 &= \vua^* \mdot \Dfxx (\vua,\vec{h}_{1100})
  + \vua^* \mdot \Dfxx (\vub,\vec{h}_{2000}) + 
  \tfrac{1}{2}\vua^* \mdot \Dfxxx(\vua,\vua,\vub),\\
  \label{eq:GExpB}
  g &= \vua^* \mdot \Dfxx (\vua,\vec{h}_{110})
  + \vua^* \mdot \Dfxx (\vub,\vec{h}_{200}) + 
  \tfrac{1}{2}\vua^* \mdot \Dfxxx(\vua,\vua,\vub),
\end{align}
which differ only by $\vec{h}_{110}$ replacing
$\vec{h}_{1100}$ and $\vec{h}_{200}$ replacing
$\vec{h}_{2000}$.  The latter pair are identical because they
are unique solutions to identical linear equations. But
$\vec{h}_{110} \neq \vec{h}_{1100}$ because $(p,q) =
(1,1,0,0)$ is resonant for the real mode in the fold-Hopf
bifurcation, whereas $(p,q) = (1,1,0)$ is nonresonant in the
Hopf bifurcation.  So the two vector coefficients are defined
through two slightly different linear equations:
\begin{align}
  \label{eq:SecTrmA}
  -\vec{J} \mdot \vec{h}_{1100} &= \Dfxx (\vua,\vub) - c_0 \vec{v},\\ 
  \label{eq:SecTrmB}
  -\vec{J} \mdot \vec{h}_{110} &= \Dfxx (\vua,\vub).
\end{align}
Here, $c_0$ is the coefficient of $\abs{w}^2$ in
\eqn{eq:DsheB}.  The second term of \eqn{eq:SecTrmA} arises
through the projection operator $\vec{Q}_3$ in
\tab{tab:FoldHopfTable}.

The physical significance is the following.  In the CGLE,
effects of components of the vector field $\vec{F}$ out of the
plane of oscillation can only be taken into account through
the amplitude transformation.  In the DSHE, the dynamics of
the slow mode is directly accounted for in the amplitude
equation.  Consequently, it does not appear in the amplitude
transformation, and as a result, the coefficients $g_1$ and
$g$ are in fact quite different.  Compare their numerical
values for the chosen operating point of the 4D Oregonator
exhibited in \tabs{tab:CGLEData}{tab:DSHEData}.  Here, it is
interesting to note that using $g_1$ in place of $g$ in the
CGLE would place the equation well inside the region of
absolute instability (see \figa{fig:Hopf4DDiff}{d}).

However, this comparison of parameters is misleading, since it
neglects the effect of the term $g_0 w z$.  To illuminate the
issue, we first find an explicit relation between
$\vec{h}_{1100}$ and $\vec{h}_{110}$ from the difference
between \eqns{eq:SecTrmA}{eq:SecTrmB},
\begin{equation}
  \vec{J} \mdot \, (\vec{h}_{1100} - \vec{h}_{110}) =
  c_0 \vec{v}.
\end{equation}

Since $\vec{J}$ is nonsingular , the solution is 
\begin{equation}
  \vec{h}_{1100} - \vec{h}_{110} =
  c_0 \vec{J}^{-1} \mdot \vec{v} =
  \frac{c_0}{\lambda_0} \vec{v}.
  \label{eq:GRel}
\end{equation}
We can then use this result to derive an explicit relation
between $g_1$ and $g$.  Solving \eqn{eq:GRel} for
$\vec{h}_{1100}$ and substituting the result in
\eqn{eq:GExpA}, we get, using the bilinearity of
$\Dfxx(\vec{u},\vec{h}_{1100})$ and the identity
$\vec{h}_{2000} = \vec{h}_{200}$:
\begin{equation}
  g_1 = g +
  \frac{c_0}{\lambda_0} \vec{u}^* \mdot 
  \Dfxx(\vec{u},\vec{v}) = g + \frac{c_0}{\lambda_0} g_0.
  \label{eq:GEq}
\end{equation}

We can get an indication of the effect of the term $g_0 wz$ by
examining the artificial situation where $z$ is constrained to
be homogeneous and stationary, so
\begin{equation}
  z = -\frac{c_0}{\lambda_0}\abs{w}^2
\end{equation}
from \eqn{eq:DsheB}.  Substitution of this result in
\eqn{eq:DsheA} then yields
\begin{equation}
  \dot{w} = \sigma_1 \mu w + 
  (g_1-\frac{c_0}{\lambda_0}) \abs{w}^2w + d_w \nabn w,
\end{equation}
which reduces to the CGLE, exactly, when \eqn{eq:GEq} and the
identity $d_w = d$ are employed.  Thus it is the variation of
$z$ with space and time---the dynamics of the slow mode---that
causes the deviation from CGLE behavior, which of course is no
surprise.  

In practice, we regain CGLE behavior if $\lambda_0$ becomes
sufficiently large and negative implying that $g_1$ and $g$
almost coalesce (according to~\eqn{eq:GEq}) and the linear
term $\lambda_0 z$ in \eqref{eq:DsheB} ensures that $\abs{z}$
and hence $\abs{g_0 w z}$ never becomes large.  The magnitude
of $\lambda_0$ should be compared with $\mu \re\{ \sigma_1\}$
first of all, and a condition for using the CGLE is
$\abs{\lambda_0} \gg \mu \re\{ \sigma_1\}$.

\newpage 
\bibliography{dynamic} 

\begin{thebibliography}{10}

\bibitem{OregOrig}
Field R.J. and Noyes R.M., 1974, Oscillations in chemical systems {IV}. {L}imit
  cycle behavior in a model of a real chemical reaction, J.\ Chem.\ Phys. 60,
  1877--1884.

\bibitem{ZhabOverview}
Zhabotinsky A.M., 1991, A history of chemical oscillations and waves, Chaos 1,
  379--386.

\bibitem{Wang95}
Wang J., Sørensen P.G. and Hynne F., 1995, Transient complex oscillations in
  the closed {Belousov-Zhabotinsky} reaction: Experimental and computational
  studies, Z.\ Phys.\ Chem. 192, 63--76.

\bibitem{Ipsen98}
Ipsen M., Hynne F. and Sørensen P.G., 1998, Systematic derivation of amplitude
  equations and normal forms for dynamical systems, Chaos 8, 834--852.

\bibitem{FKNorig}
Field R.J., Körös E. and Noyes R.M., 1972, Oscillations in chemical systems.
  {T}horough ana\-lysis of temporal oscillations in the bromate-cerium-malonic
  acid system, J.\ Amer.\ Chem.\ Soc. 94, 8649--8664.

\bibitem{FFrate}
Field R.J. and Förster\-ling H.D., 1986, On the oxybromine chemistry rate
  constants with cerium ions in the {F}ield-{K}örös-{N}oyes mechanism of the
  {B}elousov-{Z}habotinsky reaction, J.\ Phys.\ Chem. 90, 5400.

\bibitem{tina2}
Hynne F., Sørensen P.G. and Møller T., 1993, Complete optimization of models of
  the {Belousov-Zhabotinsky} reaction at a {Hopf} bifurcation, J.\ Chem.\ Phys.
  98, 219--230.

\bibitem{Ar96}
Aranson I., Levine H. and Tsimring L., 1996, Spiral competition in
  three-component excitable media, Phys.\ Rev.\ Lett. 76, 1170--1173.

\bibitem{Fitzhugh61}
FitzHugh R., 1961, Impulses and physiological states in theoretical models of
  nerve membrane, Biophys.\ J 1, 445--466.

\bibitem{Nagumo62}
Nagumo J.S., Arimoto S. and Youshizawa S., 1962, An active pulse transmission
  line simulating nerve axon, Proc.\ IRE 50, 2061--2071.

\bibitem{BohrTurb}
Bohr T., Jensen M.H., Paladin G. and Vulpiani A., 1998, Dynamical Systems
  Approach to Turbulence, Cambridge Nonlinear Science Series (Cambridge
  University Press, Cambridge).

\bibitem{Kuramoto}
Kuramoto Y., 1984, Chemical Oscillations, Waves, and Turbulence
  (Springer-Verlag, Berlin).

\bibitem{KT76}
Kuramoto Y. and Tsuzuki T., 1976, Persistent propagation of concentration waves
  in dissipative media far from thermal equilibrium, Prog.\ Theor.\ Phys. 55,
  356.

\bibitem{Si77}
Sivashinsky G.I., 1977, Nonlinear analysis of hydrodynamic instability in
  laminar flames. {Part I. D}erivation of basic equations, Acta Astronautica 4,
  1177.

\bibitem{CrossHohen}
Cross M.C. and Hohenberg P.C., 1993, Pattern formation outside of equilibrium,
  Rev.\ Mod.\ Phys. 65, 851--1112.

\bibitem{Ipsen97}
Ipsen M., Hynne F. and Sørensen P.G., 1997, Amplitude equations and chemical
  reaction-diffusion systems, Int.\ J.\ Bifurcation and Chaos 7, 1539--1554.

\bibitem{Huber92}
Huber G., Alstrøm P. and Bohr T., 1992, Nucleation and transients at the onset
  of vortex turbulence, Phys.\ Rev.\ Lett 69, 2380--2383.

\bibitem{Ar92a}
Aranson I.S., Aranson L., Kramer L. and Weber A., 1992, Stability limits of
  spirals and travelling waves in nonequilibrium media, Phys.\ Rev.\ A. 46,
  R2992--R2995.

\bibitem{Ar92b}
Aranson I.S., Golomb D. and Sompolinsky D., 1992, Spatial coherence and
  temporal chaos in macroscopic systems with asymmetrical couplings, Phys.\
  Rev.\ Lett. 68, 3495--3498.

\bibitem{MiIs98}
Ipsen M. and Schreiber I., 1998, Numerical determination of dynamical
  properties near hopf bifurcation points, preprint.

\bibitem{CgleNature}
Ouyang Q. and Flesselles J.M., 1996, Transition from spirals to defect
  turbulence driven by a convective instability, Nature 379, 143--146.

\bibitem{PgsCgle}
Hynne F. and Sørensen P.G., 1993, Experimental determination of
  {G}inzburg-{L}andau parameters for reaction-diffusion systems, Phys.\ Rev.\ E
  48, 4106--4109.

\bibitem{SW96}
Wolfram S., 1996, The Mathematica Book (Cambridge University Press, Cambridge,
  UK), third edn.

\bibitem{Maple}
Monagan M.B., Geddes K.O., Heal K.M., Labahn G. and Vorkoetter S., 1996, Maple
  V Programming Guide (Springer-Verlag, New York).

\bibitem{EnvOrg93}
Schwarzenbach R.P., Gschwend P.M. and Imboden D.H., 1993, Environmental Organic
  Chemistry (John Wiley \& Sons, Inc, New York).

\end{thebibliography}
\bibliographystyle{physica-d}

\newpage
\noindent\Large\textbf{Captions of tables and figures}
\normalsize

\vspace{2.5mm}%
\noindent\textbf{Figure~\ref{fig:Hopf4DDiff}}: 
(a): Bifurcation diagram showing the locations of
supercritical Hopf bifurcations in the 4D~Oregonator model in
the plane of the two parameters \chem{[\bromat]} and
\chem{[\hp]}. (b): Variation of the dimensionless diffusion
parameter $\alpha$ along the bifurcation curve in (a), plotted
as a function of the parameter \chem{[\bromat]}. (c):
Variation of the dimensionless parameter $\beta$ along the
branches in (a) plotted as a function of the parameter
\chem{[\bromat]}. (d): Projections of the curves in (b) and
(c) onto the $(\alpha,\beta)$-plane together with the curves
corresponding to the Eckhaus (\textsf{EH}) and absolute
instability (\textsf{AI}) curves respectively.  The
Benjamin-Feir instability line (\textsf{BF}) is also shown.

\vspace{2.5mm}%
\noindent\textbf{Figure~\ref{fig:ReacAndGinzA}}: 
Solution to the 4D~Oregonator PDE at a finite distance from
the bifurcation corresponding to an amplitude, which is 7.5\%
of the value of the stationary \chem{\cef} concentration.  The
three rows show the amplitude $\abs{w}$ and two selected
concentrations at the times shown below each column.
Integrations were made on a 256x256 grid with no-flux boundary
conditions and physical dimensions \chem{4\;cm} in both
directions.

\vspace{2.5mm}%
\noindent\textbf{Figure~\ref{fig:ReacAndGinzB}}: 
Solution to the CGLE corresponding to the one for the 4D
Oregonator shown in \fig{fig:ReacAndGinzA}. The number below
each column shows the time elapsed for that particular
state. Integrations were made on a 256x256 grid with no-flux
boundary conditions and physical dimensions \chem{4\;cm} in
both directions.  The relative variation of \chem{[BrMA]} is
extremely small (compare with \tab{tab:EigData}).  It is
represented correctly by the colors but cannot be
distinguished on the corresponding colorscale.

\vspace{2.5mm}%
\noindent\textbf{Figure~\ref{fig:Ex4DCompA}}: 
Solution to the 4D~Oregonator PDE at a finite distance from
the bifurcation point corresponding to an amplitude, 7.5\% of
the value of the stationary \chem{\cef} concentration. The
number below each column shows the time elapsed for that
particular state.  The solution is the same as in
\fig{fig:ReacAndGinzA}, but the amplitude $z$ is exhibited
instead of $\abs{w}$.  Integrations were made on a 256x256
grid with no-flux boundary conditions and physical dimensions
\chem{4\;cm} in both directions.

\vspace{2.5mm}%
\noindent\textbf{Figure~\ref{fig:Ex4DCompB}}: 
Solution to the DSHE corresponding to that for the 4D
Oregonator shown in \fig{fig:Ex4DCompA}. The number below each
column shows the time elapsed for that particular
state. Integrations were made on a 256x256 grid with no-flux
boundary conditions and physical dimensions \chem{4\;cm} in
both directions.

\vspace{2.5mm}%
\noindent\textbf{Figure~\ref{fig:SteadyTrans}}: 
Time series showing the local variation of $\re w$ (top) and
$z$ (bottom) at a particular grid point in the course of the
pattern development shown in \fig{fig:Ex4DCompB}. Observe that
the time variation of $z$ is much slower than for $\re w$.

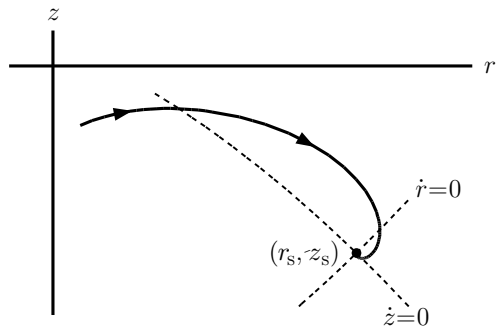
\begin{figure}[!t]
  \footnotesize
  \begin{center}
    \leavevmode
    \input{Figures/fig9.tex}
    \caption{%
      }%
    \label{fig:NullCl}
  \end{center}
  \normalsize
\end{figure}

\vspace{2.5mm}%
\noindent\textbf{Figure~\ref{fig:NullCl}}: 
(a): Solution to \eqn{eq:DsheAmp2} approaching the stable
focus $(\stat{r},\stat{z})$.

\vspace{2.5mm}%
\noindent\textbf{Table~\ref{tab:FoldHopfTable}}: 
Formulas for calculating the coefficients of the amplitude
transformation and amplitude equation for the fold-Hopf
bifurcation.  At the bifurcation, the Jacobian $\vec{J}$ has
three eigenvectors $\vec{u}$, $\cc{\vec{u}}$, and $\vec{v}$
and left eigenvectors $\vec{u}^*$, $\cc{\vec{u}}^*$, and
$\vec{v}^*$ corresponding to the three critical eigenvalues
$\lambda_1 = \cc{\lambda}_2 = \I\omega_0$ and $\lambda_3=0$.
The amplitude transformation $\vec{x} = \vec{y} +
\vec{h}(\vec{y},\mu)$, $\vec{y} = \vec{u} w \E^{\I \omega_0 t}
+ \mathrm{c.c.} + \vec{v}z$, transforms a solution
$w(\vec{r},t),z(\vec{r},t)$ of the amplitude equation to the
motion $\vec{x}(\vec{r},t)$ on the unfolded center manifold
for the dynamical system.  The vector coefficients
$\vec{h}_{pqrs}$ are determined as solutions to the linear
equations indicated, in terms of the derivatives of the vector
field $\vec{F}$.  The coefficients of the amplitude equation
can then be found through the explicit expressions indicated,
in terms of the derivatives of $\vec{F}$ and $\vec{h}_{pqrs}$.
For any $\vec{x} \in \R^n$, the projcetions $\vec{Q}_1$ and
$\vec{Q}_3$ are defined as $\vec{Q}_1 \mdot \vec{x} = \vec{x}
- (\vec{u}^* \mdot \vec{x})\vec{u}$ and $\vec{Q}_3 \mdot
\vec{x} = \vec{x} - (\vec{v}^* \mdot \vec{x})\vec{v}$
respectively.

\vspace{2.5mm}%
\noindent\textbf{Table~\ref{tab:HopfTable}}: 
Formulas for calculating the coefficients of the amplitude
transformation and amplitude equation for the Hopf
bifurcation.  At the bifurcation, the Jacobian $\vec{J}$ has
two complex conjugate eigenvectors $\vec{u}$ and
$\cc{\vec{u}}$ and left eigenvectors $\vec{u}^*$ and
$\cc{\vec{u}}^*$ corresponding to critical eigenvalues
$\lambda_1 = \cc{\lambda}_2 = \I\omega_0$.  The amplitude
transformation $\vec{x} = \vec{y} + \vec{h}(\vec{y},\mu)$,
$\vec{y} = \vec{u}w\E^{\I \omega_0 t} + \mathrm{c.c.}$,
transforms a solution $w(\vec{r},t)$ of the amplitude equation
to the motion $\vec{x}(\vec{r},t)$ on the unfolded center
manifold for the dynamical system.  The vector coefficients
$\vec{h}_{pqs}$ are determined as solutions to the linear
equations indicated, in terms of the derivatives of the vector
field~$\vec{F}$.  The coefficients of the amplitude equation
can then be found through the explicit expressions indicated,
in terms of the derivatives of $\vec{F}$ and $\vec{h}_{pqs}$.
For any $\vec{x} \in \R^n$, the projcetion $\vec{Q}_1$ is
defined as $\vec{Q}_1 \mdot \vec{x} = \vec{x} - (\vec{u}^*
\mdot \vec{x})\vec{u}$.

\vspace{2.5mm}%
\noindent\textbf{Table~\ref{tab:Diff4DOrg}}: 
Rate constants $k_1,\ldots,k_7$, constant malonic acid
concentration $M$, and diffusion constants $D_X$, $D_Y$,
$D_Z$, and $D_U$ used for numerical simulations for the 4D
Oregonator reaction-diffusion equation~\eno{eq:Diff4DOrg}. The
values of the rate constants are based on \cite{Wang95} but
have been modified slightly in the calculations presented
here. Diffusion constants $D_X,\ldots,D_Z$ are from
\cite{PgsCgle}, whereas the value of $D_U$ was estimated by an
interpolation scheme described in \protect\cite[p.\
196]{EnvOrg93}.

\vspace{2.5mm}%
\noindent\textbf{Table~\ref{tab:HopfData}}:  
Hopf bifurcation point for the 4D Oregonator used as a
reference for the operating point actually used in the
calculations. The table shows the parameters, the stationary
point, and the eigenvalues $\pm$i$\omega_0$, $\lambda_3$, and
$\lambda_4$ of the Jacobian matrix at that point.

\vspace{2.5mm}%
\noindent\textbf{Table~\ref{tab:CGLEData}}:  
Ginzburg-Landau parameters at the Hopf bifurcation,
\tab{tab:HopfData}, for the 4D Oregonator \eqn{eq:Diff4DOrg}
with parameters shown in \tab{tab:Diff4DOrg}.

\vspace{2.5mm}%
\noindent\textbf{Table~\ref{tab:DSHEData}}: 
Coefficients for the distributed slow-Hopf
equation~\eno{eq:DSHE} at the Hopf bifurcation,
\tab{tab:HopfData}, for the 4D Oregonator \eqn{eq:Diff4DOrg}
with parameters shown in \tab{tab:Diff4DOrg}.

\vspace{2.5mm}%
\noindent\textbf{Table~\ref{tab:EigData}}: 
Left and right eigenvectors of the Jacobian matrix at the Hopf
bifurcation, \tab{tab:HopfData}, for the 4D Oregonator
\eqn{eq:Diff4DOrg} with parameters shown in
\tab{tab:Diff4DOrg}.  The eigenvectors $\vua^*$ (left) and
$\vuc$ (right) correspond to eigenvalues $\I\omega_0$ and
$\lambda_3$ respectively.  They illuminate the special role
played by BrMA in the 4D Oregonator.

\renewcommand{\baselinestretch}{1.00}

\begin{sidewaysfigure}
  \leavevmode
  \begin{center}
    \begin{pspicture}(0,0)(\textheight,14.5)
  \sffamily
  \psframe(0,0)(\textheight,14.5)

  \rput[cc]( 9.0,13.3){(a)}
  \rput[cc](19.5,13.3){(b)}
  \rput[cc]( 9.0, 6.2){(c)}
  \rput[cc](19.5, 6.2){(d)}

  \rput[rc](13.3,6.1){\scriptsize EH}
  \rput[lc](13.9,5.3){\scriptsize AI}
  \rput[bc](18.0,5.5){\scriptsize BF}

  \rput[lc]( 5.0,10.8){
\setlength{\unitlength}{0.1bp}
\begin{picture}(2700,1943)(0,0)
\special{psfile=Figures/fig1a.ps llx=0 lly=0 urx=540 ury=454 rwi=5400}
\put(1233,1699){\makebox(0,0){\BlckBox}}
\put(2700,568){\makebox(0,0){\CircBox{\tiny\textsf{3}}}}
\put(2700,484){\makebox(0,0){\CircBox{\tiny\textsf{2}}}}
\put(768,1843){\makebox(0,0){\CircBox{\tiny\textsf{1}}}}
\put(1600,50){\makebox(0,0){\large \chem{\sf [BrO_3^-]/\molarsf}}}
\put(261,1071){%
\special{ps: gsave currentpoint currentpoint translate
270 rotate neg exch neg exch translate}%
\makebox(0,0)[b]{\shortstack{\large {\chem{\sf [H^+]/\molarsf}}}}%
\special{ps: currentpoint grestore moveto}%
}
\put(2312,200){\makebox(0,0){1.2}}
\put(1924,200){\makebox(0,0){0.9}}
\put(1535,200){\makebox(0,0){0.6}}
\put(1147,200){\makebox(0,0){0.3}}
\put(759,200){\makebox(0,0){0}}
\put(450,1571){\makebox(0,0)[r]{1.2}}
\put(450,1298){\makebox(0,0)[r]{0.9}}
\put(450,1026){\makebox(0,0)[r]{0.6}}
\put(450,754){\makebox(0,0)[r]{0.3}}
\put(450,482){\makebox(0,0)[r]{0}}
\end{picture}}
  \rput[lc](15.4,10.8){
\setlength{\unitlength}{0.1bp}
\begin{picture}(2700,1943)(0,0)
\special{psfile=Figures/fig1b.ps llx=0 lly=0 urx=540 ury=454 rwi=5400}
\put(1037,447){\makebox(0,0){\BlckBox}}
\put(2700,824){\makebox(0,0){\CircBox{\tiny\textsf{3}}}}
\put(2700,1187){\makebox(0,0){\CircBox{\tiny\textsf{2}}}}
\put(584,1843){\makebox(0,0){\CircBox{\tiny\textsf{1}}}}
\put(1600,50){\makebox(0,0){\large \chem{\sf [BrO_3^-]/\molarsf}}}
\put(300,1071){\makebox(0,0)[b]{\shortstack{\large $\alpha$}}}
\put(2260,200){\makebox(0,0){1.2}}
\put(1820,200){\makebox(0,0){0.9}}
\put(1380,200){\makebox(0,0){0.6}}
\put(940,200){\makebox(0,0){0.3}}
\put(450,1534){\makebox(0,0)[r]{8}}
\put(450,1226){\makebox(0,0)[r]{6}}
\put(450,917){\makebox(0,0)[r]{4}}
\put(450,609){\makebox(0,0)[r]{2}}
\end{picture}}
  \rput[lc]( 5.0, 3.7){
\setlength{\unitlength}{0.1bp}
\begin{picture}(2700,1943)(0,0)
\special{psfile=Figures/fig1c.ps llx=0 lly=0 urx=540 ury=454 rwi=5400}
\put(1037,1561){\makebox(0,0){\BlckBox}}
\put(2700,801){\makebox(0,0){\CircBox{\tiny\textsf{3}}}}
\put(2700,1050){\makebox(0,0){\CircBox{\tiny\textsf{2}}}}
\put(525,529){\makebox(0,0){\CircBox{\tiny\textsf{1}}}}
\put(1600,50){\makebox(0,0){\large \chem{\sf [BrO_3^-]/\molarsf}}}
\put(150,1071){\makebox(0,0)[b]{\shortstack{\large $\beta$}}}
\put(2260,200){\makebox(0,0){1.2}}
\put(1820,200){\makebox(0,0){0.9}}
\put(1380,200){\makebox(0,0){0.6}}
\put(940,200){\makebox(0,0){0.3}}
\put(450,1586){\makebox(0,0)[r]{-0.2}}
\put(450,1072){\makebox(0,0)[r]{-0.4}}
\put(450,557){\makebox(0,0)[r]{-0.6}}
\end{picture}}
  \rput[lc](15.4, 3.7){
\setlength{\unitlength}{0.1bp}
\begin{picture}(2700,1943)(0,0)
\special{psfile=Figures/fig1d.ps llx=0 lly=0 urx=540 ury=454 rwi=5400}
\put(763,1303){\makebox(0,0){\BlckBox}}
\put(1435,544){\makebox(0,0){\CircBox{\tiny\textsf{3}}}}
\put(2080,792){\makebox(0,0){\CircBox{\tiny\textsf{2}}}}
\put(2700,467){\makebox(0,0){\CircBox{\tiny\textsf{1}}}}
\put(1600,50){\makebox(0,0){\large $\alpha$}}
\put(150,1071){\makebox(0,0)[b]{\shortstack{\large $\beta$}}}
\put(2425,200){\makebox(0,0){7}}
\put(1875,200){\makebox(0,0){5}}
\put(1325,200){\makebox(0,0){3}}
\put(775,200){\makebox(0,0){1}}
\put(450,1586){\makebox(0,0)[r]{-0.1}}
\put(450,1329){\makebox(0,0)[r]{-0.2}}
\put(450,1071){\makebox(0,0)[r]{-0.3}}
\put(450,814){\makebox(0,0)[r]{-0.4}}
\put(450,557){\makebox(0,0)[r]{-0.5}}
\end{picture}}
\end{pspicture}

  \end{center}
  \caption{%
    }
  \label{fig:Hopf4DDiff}
\end{sidewaysfigure}

\begin{sidewaysfigure}[htbp]
  \begin{center}
    \leavevmode
    \caption{%
      }
    \label{fig:ReacAndGinzA}
  \end{center}
\end{sidewaysfigure}

\begin{sidewaysfigure}[htbp]
  \begin{center}
    \leavevmode
    \caption{%
      }%
    \label{fig:ReacAndGinzB}
  \end{center}
\end{sidewaysfigure}

\begin{sidewaysfigure}[htbp]
  \begin{center}
    \leavevmode
    \caption{%
      }%
    \label{fig:Ex4DCompA}
  \end{center}
\end{sidewaysfigure}

\begin{sidewaysfigure}[htbp]
  \begin{center}
    \leavevmode
    \caption{%
      }%
    \label{fig:Ex4DCompB}
  \end{center}
\end{sidewaysfigure}

\begin{figure}[!t]
  \begin{center}
    \leavevmode
    \caption{
      }%
    \label{fig:SteadyTrans}
  \end{center}
  \normalsize
\end{figure}

\clearpage
\begin{table}[t!]
  \begin{center}
    \ScaleXSize{%
      \newlength{\AlgSpace}
\newlength{\TopSpace}
\newlength{\IntSpace}
\newcommand{\PushSpace}{\centering}
\setlength{\AlgSpace} {0.9mm}
\setlength{\TopSpace} {2.0mm}
\setlength{\IntSpace} {1.0mm}
\footnotesize
\begin{tabular}{%
    |>{\footnotesize} m{2.0cm} <{}
    |>{\footnotesize\vspace{\TopSpace}} m{7.9cm} <{\vspace{\TopSpace}}
     >{\footnotesize\vspace{\TopSpace}} m{1.80cm} <{\vspace{\TopSpace}}
     >{\footnotesize\vspace{\TopSpace}} m{1.90cm} <{\vspace{\TopSpace}}
    |}\hline
\multicolumn{4}{|c|}{\emph{\footnotesize Fold-Hopf Bifurcation}} 
\\\hline\hline

\textsf{Transformation}\newline
$\vec{x} = \vec{y} + \vec{h}(\vec{y},\mu)$ &
\multicolumn{3}{>{\footnotesize\vspace{\TopSpace}} m{12.0cm} <{\vspace{\TopSpace}}|}{%
  $\begin{aligned}      
    \vec{x} &=  
    \vua \psi + \cc{\vua\psi} + \vuc z +
    \vec{h}_{2000}\psi^2 + \vec{h}_{1100}\abs{\psi}^2  + \vec{h}_{0200}\cc{\psi}^2 +
    \vec{h}_{0020}z^2 + \vec{h}_{1010}\psi z + \vec{h}_{0110}
    \cc{\psi}z +\\
    &
    \vec{h}_{3000}\psi^3 + \vec{h}_{2100}\abs{\psi}^2\psi +
    \vec{h}_{1200}\abs{\psi}^2 \cc{\psi} + \vec{h}_{0300}\cc{\psi}^3  +
    \vec{h}_{0030}z^3 + \vec{h}_{2010}\psi^2z + 
    \vec{h}_{1110}\abs{\psi}^2z + \\
    &
    \vec{h}_{0210}\cc{\psi}^2 z + 
    \vec{h}_{1020}\psi z^2 + \vec{h}_{0120}\cc{\psi} z^2 + 
    (\vec{h}_{0001} + \vec{h}_{1001} \psi + \vec{h}_{0101} \cc{\psi}
    + \vec{h}_{0011} z)\mu, 
    \qquad\quad \psi = w\E^{\I \omega_0 t}
  \end{aligned}$
}
\\\hline
\multicolumn{4}{c}{} 
\\[\IntSpace]\hline

\multicolumn{1}{|c|}{\textsf{Second order}} & 
\multicolumn{3}{c|}{\emph{Linear equations for $\vec{h}_{pqrs}$}}\\\hline

\PushSpace$w^2$ &
$\begin{eqalign}
  -(\vec{J} - \ome{2}) \mdot \vec{h}_{2000} &= 
  \tfrac{1}{2}\Dfxx (\vua,\vua)
\end{eqalign}$ 
& 
& $\vec{h}_{0200} = \cc{\vec{h}}_{2000}$ \\\hline

\PushSpace$\abs{w}^2$ &
$\begin{eqalign}
  -\vec{J} \mdot \vec{h}_{1100} &= 
  -\vec{Q}_3 \mdot \Dfxx (\vua,\vub)
\end{eqalign}$ & & \\\hline

\PushSpace$w z$ &
$\begin{eqalign}
  -(\vec{J} - \ome{}) \mdot \vec{h}_{1010} &= 
  \vec{Q}_1 \mdot \Dfxx (\vua,\vuc)
\end{eqalign}$ 
& $\vua^* \mdot \vec{h}_{0020} = 0$
& $\vec{h}_{0110} = \cc{\vec{h}}_{1010}$ \\\hline

\PushSpace$z^2$ &
$\begin{eqalign}
  -\vec{J} \mdot \vec{h}_{0020} &= 
  \tfrac{1}{2} \vec{Q}_3 \mdot \Dfxx (\vuc,\vuc)
\end{eqalign}$ 
& $\vuc^* \mdot \vec{h}_{0020} = 0$
& \\\hline

\multicolumn{4}{c}{} 
\\[\IntSpace]\hline

\multicolumn{1}{|c|}{\textsf{Third order}} & 
\multicolumn{3}{c|}{\emph{Linear equations for $\vec{h}_{pqrs}$}}\\\hline\hline

\PushSpace$w^3$ &
$\begin{eqalign}
  -(\vec{J} - \ome{3}) \mdot 
  \vec{h}_{3000} &= \Dfxx (\vua,\vec{h}_{2000}) +
  \tfrac{1}{6}\Dfxxx (\vua,\vua, \vua)
\end{eqalign}$ 
&
& $\vec{h}_{0300} = \cc{\vec{h}}_{3000}$ \\\hline

\PushSpace$\abs{w}^2w$ &
$\begin{eqalign}
  -(\vec{J} - \ome{}) \mdot \vec{h}_{2100} = 
  \vec{Q}_1 \mdot \bigl( 
     \Dfxx (\vua,\vec{h}_{1100}) + \\
     \qquad\qquad
     \Dfxx (\vub,\vec{h}_{2000}) + 
     \tfrac{1}{2}\Dfxxx (\vua,\vua,\vub) \bigl)
\end{eqalign}$ 
& $\vua^* \mdot \vec{h}_{2100} = 0$
& $\vec{h}_{1200} = \cc{\vec{h}}_{2100}$ \\\hline

\PushSpace$w z^2$ &
$\begin{eqalign}
  -(\vec{J} - \ome{}) \mdot \vec{h}_{1020} = 
  \vec{Q}_1 \mdot\, (
    \Dfxx  (\vua,\vec{h}_{0020}) + \\
    \qquad\qquad
    \Dfxx  (\vuc,\vec{h}_{1010}) +
  \tfrac{1}{2}\Dfxxx (\vua,\vuc,\vuc) )
\end{eqalign}$ 
&
& $\vec{h}_{0120} = \cc{\vec{h}}_{1020}$ \\\hline

\PushSpace$\abs{w}^2 z$ &
\multicolumn{3}{>{\footnotesize\vspace{\TopSpace}} m{12.0cm} <{\vspace{\TopSpace}}|}{%
$\begin{eqalign}
  -\vec{J} \mdot \vec{h}_{1110} &= 
  \vec{Q}_3 \mdot\, (
   \Dfxx  (\vua,\vec{h}_{0110}) +
   \Dfxx  (\vub,\vec{h}_{1010}) +
   \Dfxx  (\vuc,\vec{h}_{1100}) +
   \Dfxxx (\vua,\vub,\vuc) )
\end{eqalign}$ 
} \\\hline

\PushSpace$w^2 z$ &
$\begin{eqalign}
  -(\vec{J} - \ome{2}) \mdot \vec{h}_{2010} = 
    \Dfxx (\vua,\vec{h}_{1010}) + 
    \Dfxx (\vuc,\vec{h}_{2000}) + \\
    \qquad\qquad
    \tfrac{1}{2}\Dfxxx (\vua,\vua,\vuc) - g_2 \vec{h}_{2000}
\end{eqalign}$  
&
& $\vec{h}_{0210} = \cc{\vec{h}}_{2010}$ \\\hline

\PushSpace$z^3$ &
$\begin{eqalign}
  -\vec{J} \mdot \vec{h}_{0030} &= 
   \vec{Q}_3 \mdot\, (
                \Dfxx  (\vuc,\vec{h}_{0020}) +
   \tfrac{1}{6} \Dfxxx (\vuc,\vuc,\vuc))
\end{eqalign}$
& $\vuc^* \mdot \vec{h}_{0030} = 0$
& \\\hline

\multicolumn{4}{c}{} 
\\[\IntSpace]\hline

\multicolumn{1}{|c|}{\textsf{Unfoldings}} & 
\multicolumn{3}{c|}{\emph{Linear equations for $\vec{h}_{pqrs}$}}\\\hline\hline

\PushSpace$\mu$ &
$\begin{eqalign}
  -\vec{J} \mdot \vec{h}_{0001} &= \vec{Q}_3 \mdot \Dfp
\end{eqalign}$ 
& $\vuc^* \mdot \vec{h}_{0001} = 0$
& \\\hline

\PushSpace$\mu w$ &
$\begin{eqalign}
  -(\vec{J} - \ome{}) \mdot \vec{h}_{1001} &= 
  \vec{Q}_1 \mdot \bigl(
    \Dfxp \mdot \vua + 
    \Dfxx (\vua,\vec{h}_{0001}) \bigr)
\end{eqalign}$ 
& $\vua^* \mdot \vec{h}_{1001} = 0$
& $\vec{h}_{0101} = \cc{\vec{h}}_{1001}$ \\\hline

\PushSpace$\mu z$ &
$\begin{eqalign}
  -\vec{J} \mdot \vec{h}_{0011} &= 
  \vec{Q}_3 \mdot \bigl(
    \Dfxp \mdot \vuc + 
    \Dfxx (\vuc,\vec{h}_{0001}) \bigr)
\end{eqalign}$ 
& $\vuc^* \mdot \vec{h}_{0011} = 0$ & \\\hline

\multicolumn{4}{c}{} 
\\[\IntSpace]\hline

\textsf{Amplitude}\newline 
\textsf{equation} 
& 
$\begin{eqalign}
  \dot{w} &= \sigma_1 \mu w + g_0 w z + g_1 \abs{w}^2w + g_2 w z^2 +
  d_w \nabn w,\\
  \dot{z} &= \rho_0 \mu + c_0 \abs{w}^2 + c_1 z^2 + 
  c_2 \abs{w}^2 z + c_3 z^3 + d_z \nabn z.
\end{eqalign}$ 
& 
\multicolumn{2}{c|}{%
  $d_w = \vua^* \mdot \vec{D} \mdot \vua, \quad
   d_z = \vuc^* \mdot \vec{D} \mdot \vuc$} \\\hline

\textsf{Resonant}\newline
\textsf{coefficients} & \multicolumn{3}{>{\footnotesize\vspace{\TopSpace}} m{12.0cm} <{\vspace{\TopSpace}}|}{%
  $\begin{eqalign}
    g_0 &= \vua^* \mdot \Dfxx (\vua,\vuc)
    \\[\AlgSpace]
    g_1 &= \vua^* \mdot \Dfxx (\vua,\vec{h}_{1100})
    + \vua^* \mdot \Dfxx (\vub,\vec{h}_{2000}) + 
    \tfrac{1}{2}\vua^* \mdot \Dfxxx(\vua,\vua,\vub)
    \\[\AlgSpace]
    g_2 &= 
    \vua^* \mdot \Dfxx (\vua,\vec{h}_{0020}) +
    \vua^* \mdot \Dfxx (\vuc,\vec{h}_{1010}) +
    \tfrac{1}{2}\vua^* \mdot \Dfxxx(\vua,\vuc,\vuc)
    \\[\AlgSpace]
    c_0 &= \vuc^* \mdot \Dfxx (\vua,\vub), \qquad
    c_1 = \tfrac{1}{2}\vuc^* \mdot \Dfxx (\vuc,\vuc)
    \\[\AlgSpace]
    c_2 &= 
    \vuc^* \mdot \Dfxx (\vua,\vec{h}_{0110}) +
    \vuc^* \mdot \Dfxx (\vub,\vec{h}_{1010}) +
    \vuc^* \mdot \Dfxx (\vuc,\vec{h}_{1100}) +
    \vuc^* \mdot \Dfxxx (\vua,\vub,\vuc) 
    \\[\AlgSpace]
    c_3 &= 
    \vuc^* \mdot \Dfxx (\vuc,\vec{h}_{0020}) +
    \tfrac{1}{6}\vuc^* \mdot \Dfxxx (\vuc,\vuc,\vuc) 
    \\[\AlgSpace]
    \sigma_1 &= \vua^* \mdot \Dfxp \mdot \vua + 
    \vua^* \mdot \Dfxx (\vua,\vec{h}_{0001}), \qquad
    \rho_0 = \vuc^* \mdot \Dfp
  \end{eqalign}$
  }
\\\hline

\end{tabular}


      }
    \caption{%
      }%
    \label{tab:FoldHopfTable}
  \end{center}
\end{table}

\clearpage
\begin{table}[t!]
  \begin{center}
    \ScaleXSize{%
      \newcommand{\PushSpace}{\centering}
\setlength{\AlgSpace} {1.0mm}
\setlength{\TopSpace} {1.9mm}
\setlength{\IntSpace} {2.0mm}
\footnotesize
\begin{tabular}{%
    |>{\footnotesize} m{2.0cm} <{}
    |>{\footnotesize\vspace{\TopSpace}} m{7.9cm} <{\vspace{\TopSpace}}
     >{\footnotesize\vspace{\TopSpace}} m{1.80cm} <{\vspace{\TopSpace}}
     >{\footnotesize\vspace{\TopSpace}} m{1.90cm} <{\vspace{\TopSpace}}
    |}\hline
\multicolumn{4}{|c|}{\emph{\footnotesize Hopf Bifurcation}} 
\\\hline\hline

\textsf{Transformation}\newline
$\vec{x} = \vec{y} + \vec{h}(\vec{y},\mu)$ &
\multicolumn{3}{>{\footnotesize\vspace{\TopSpace}} m{12.0cm} <{\vspace{\TopSpace}}|}{%
  $\begin{aligned}      
    \vec{x} &=  
    \vua \psi + \vub \cc{\psi} +
    \vec{h}_{200}\psi^2 + \vec{h}_{110}\abs{\psi}^2  + 
    \vec{h}_{020}\cc{\psi}^2 +
    \vec{h}_{300}\psi^3 + \vec{h}_{210}\abs{\psi}^2\psi +\\
    &\qquad\vec{h}_{120}\abs{\psi}^2 \cc{\psi} + \vec{h}_{030}\cc{\psi}^3  +
    \vec{h}_{001}\mu + (\vec{h}_{101}\psi +
    \vec{h}_{011}\cc{\psi})\mu,
    \,\qquad\qquad\qquad\qquad\qquad \psi = w\E^{\I \omega_0 t}
  \end{aligned}$
}
\\\hline
\multicolumn{4}{c}{} 
\\[\IntSpace]\hline

\multicolumn{1}{|c|}{\textsf{Second order}} & 
\multicolumn{3}{c|}{\emph{Linear equations for $\vec{h}_{pqs}$}}\\\hline\hline

\PushSpace$w^2$ &
$\begin{eqalign}
  -(\vec{J} - \ome{2}) \mdot \vec{h}_{200} &= 
  \tfrac{1}{2}\Dfxx (\vua,\vua)
\end{eqalign}$ 
& 
& $\vec{h}_{020} = \cc{\vec{h}}_{200}$ \\\hline

\PushSpace$\abs{w}^2$ &
$\begin{eqalign}
  -\vec{J} \mdot \vec{h}_{110} &= \Dfxx (\vua,\vub)
\end{eqalign}$ & & \\\hline

\multicolumn{4}{c}{} 
\\[\IntSpace]\hline

\multicolumn{1}{|c|}{\textsf{third order}} & 
\multicolumn{3}{c|}{\emph{Linear equations for $\vec{h}_{pqs}$}}\\\hline\hline

\PushSpace$w^3$ &
$\begin{eqalign}
  -(\vec{J} - \ome{3}) \mdot 
  \vec{h}_{300} &= \Dfxx (\vua,\vec{h}_{200}) +
  \tfrac{1}{6}\Dfxxx (\vua,\vua, \vua)
\end{eqalign}$ 
&
& $\vec{h}_{030} = \cc{\vec{h}}_{300}$ \\\hline

\PushSpace$\abs{w}^2w$ &
$\begin{eqalign}
  -(\vec{J} - \ome{}) \mdot \vec{h}_{210} = 
  \vec{Q}_1 \mdot \bigl( 
     \Dfxx (\vua,\vec{h}_{110}) + \\
     \qquad\qquad
     \Dfxx (\vub,\vec{h}_{200}) + 
     \tfrac{1}{2}\Dfxxx (\vua,\vua,\vub) \bigl)
\end{eqalign}$ 
& $\vua^* \mdot \vec{h}_{210} = 0$
& $\vec{h}_{120} = \cc{\vec{h}}_{210}$ \\\hline

\multicolumn{4}{c}{} 
\\[\IntSpace]\hline

\multicolumn{1}{|c|}{\textsf{Unfoldings}} & 
\multicolumn{3}{c|}{\emph{Linear equations for $\vec{h}_{pqs}$}}\\\hline\hline

\PushSpace$\mu$ &
$\begin{eqalign}
  -\vec{J} \mdot \vec{h}_{001} &= \Dfp
\end{eqalign}$ 
& 
& \\\hline

\PushSpace$\mu w$ &
$\begin{eqalign}
  -(\vec{J} - \ome{}) \mdot \vec{h}_{101} &= 
  \vec{Q}_1 \mdot\, \bigl(
    \Dfxp \mdot \vua + 
    \Dfxx (\vua,\vec{h}_{001}) \bigr)
\end{eqalign}$ 
& $\vua^* \mdot \vec{h}_{101} = 0$
& $\vec{h}_{011} = \cc{\vec{h}}_{101}$ \\\hline

\multicolumn{4}{c}{} 
\\[\IntSpace]\hline

\textsf{Amplitude}\newline 
\textsf{equation} 
& 
$\begin{eqalign}
  \dot{w} &= \sigma_1 \mu w + g \abs{w}^2w + d \nabn w
\end{eqalign}$ 
& 
\multicolumn{2}{l|}{%
  $d = \vua^* \mdot \vec{D} \mdot \vua$}
 \\\hline

\textsf{Resonant}\newline
\textsf{coefficients} & \multicolumn{3}{>{\footnotesize\vspace{\TopSpace}} m{12.0cm} <{\vspace{\TopSpace}}|}{%
  $\begin{eqalign}
    g &= \vua^* \mdot \Dfxx (\vua,\vec{h}_{110})
    + \vua^* \mdot \Dfxx (\vub,\vec{h}_{200}) + 
    \tfrac{1}{2}\vua^* \mdot \Dfxxx(\vua,\vua,\vub)
    \\[\AlgSpace]
    \sigma_1 &= \vua^* \mdot \Dfxp \mdot \vua + 
    \vua^* \mdot \Dfxx (\vua,\vec{h}_{001})
  \end{eqalign}$
  }
\\\hline

\end{tabular}


      }
    \caption{%
      }%
    \label{tab:HopfTable}
  \end{center}
\end{table}

\begin{table}[htbp]
  \begin{center}
    \begin{tabular}{|>{$}l<{$}|d{6}|}\hline
  \emph{Constant} & 
  \multicolumn{1}{c|}{\emph{Value}}           \\\hline\hline
  k_1/\chem{\molar^{-3}s^{-1}}  & 1.6         \\ 
  k_2/\chem{\molar^{-2}s^{-1}}  & 2.5 \sci{6} \\ 
  k_3/\chem{\molar^{-2}s^{-1}}  & 33.0        \\ 
  k_4/\chem{\molar^{-1}s^{-1}}  & 3.0 \sci{3} \\ 
  k_5/\chem{\molar^{-1}s^{-1}}  & 30.0        \\ 
  k_6/\chem{\molar^{-2}s^{-1}}  & 0.18        \\ \hline
\end{tabular}
\hspace{1.0cm}
\begin{tabular}{|>{$}l<{$}|d{8}|}         \hline
  \emph{Constant} & 
  \multicolumn{1}{c|}{\emph{Value}}     \\\hline\hline
  k_7/\chem{\molar^{-1}}& 0.0003        \\ 
  M/\molar              & 0.44          \\ 
  D_X/\chem{cm^2s^{-1}} & 1.0 \sci{-5}  \\ 
  D_Y/\chem{cm^2s^{-1}} & 1.6 \sci{-5}  \\ 
  D_Z/\chem{cm^2s^{-1}} & 0.6 \sci{-5}  \\
  D_U/\chem{cm^2s^{-1}} & 6.7 \sci{-6}  \\ \hline
\end{tabular}

    \caption{%
      }
    \label{tab:Diff4DOrg}
  \end{center}
\end{table}

\begin{table}[htbp]
  \begin{center}
\begin{tabular}{|>{$}l<{$} d{10}|}\hline
  \emph{Parameter} & 
  \multicolumn{1}{c|}{\emph{Value}}                \\\hline\hline
  \chem{[\bromat]_{Hopf}/\molar}        &   0.3662 \\ 
  \chem{[\hp]_{Hopf}/\molar}            &   1.3416 \\ 
  \chem{[\hbro]_0/\molar}               &   9.8679\sci{-7} \\ 
  \chem{[\bromid]_0/\molar}             &   7.0929\sci{-6} \\ 
  \chem{[\cef]_0/\molar}                &   1.3185\sci{-5} \\ 
  \chem{[BrMA]_0/\molar}                &   7.8262\sci{-2} \\
  \omega_0/\chem{s^{-1}}                &   2.1911 \\
  \lambda_3/\chem{s^{-1}}               &  -3.0731\sci{-4}           \\
  \lambda_4/\chem{s^{-1}}               & -10.4379                  \\\hline
\end{tabular}

    \caption{%
      }
    \label{tab:HopfData}
  \end{center}
\end{table}

\begin{table}[htbp]
  \begin{center}
\begin{tabular}{|>{$}l<{$} d{14}|}\hline
  \emph{Parameter} & 
  \multicolumn{1}{c|}{\emph{Value}}                \\\hline\hline
  \sigma_1/\chem{s^{-1}}              &-2.57 \imag{+}{1.14}\\
  g/\chem{10^{10}\,\molar^{-2}s^{-1}} & 2.19 \imag{+}{2.09}   \\
  d/\chem{10^{-5}\,cm^2s^{-1}}        & 1.00 \imag{-}{2.10}   \\
  \alpha                              & 0.96                  \\
  \beta                               &-0.21                  \\\hline
\end{tabular}
  
    \caption{%
      }
    \label{tab:CGLEData}
  \end{center}
\end{table}

\begin{table}[htbp]
  \begin{center}
      \begin{tabular}{|>{$}l<{$} d{14}|}\hline
    \emph{Parameter} & 
    \multicolumn{1}{c|}{\emph{Value}}                     \\\hline\hline
    \lambda_0/\chem{s^{-1}}               &-3.07 \sci{-4} \\
    \sigma_1/\chem{s^{-1}}                &-2.57 \imag{+}{1.14}\\
    g_0/\chem{10^{ 5}\,\molar^{-1}s^{-1}} & 0.64 \imag{+}{1.00} \\
    g_1/\chem{10^{10}\,\molar^{-2}s^{-1}} & 1.09 \imag{+}{3.80} \\
    c_0/\chem{10\,\molar^{-1}s^{-1}}      &-5.25                \\
    d_w/\chem{10^{-5}\,cm^2s^{-1}}        & 1.00 \imag{-}{2.10} \\
    d_z/\chem{10^{-6}\,cm^2s^{-1}}        & 6.70                \\\hline
  \end{tabular}

    \caption{%
      }
    \label{tab:DSHEData}
  \end{center}
\end{table}

\begin{table}[htbp]
  \begin{center}
      \begin{tabular}{| l | d{10} | d{10} | d{10} |}\hline
    \emph{Species} &
    \multicolumn{1}{ c|}{$\re\vec{u}^*$} & 
    \multicolumn{1} {c|}{$\im\vec{u}^*$} &
    \multicolumn{1} {c|}{$\vec{v}$}       \\\hline\hline
    \chem{\brsyr}   & 7.546\sci{-1}  &  -1.843          &  3.382\sci{-3} \\ 
    \chem{\bromid}  & 1.699\sci{-1}  &   8.390\sci{-1}  & -1.138\sci{-2} \\ 
    \chem{\cef}     & 4.665\sci{-1}  &   2.519\sci{-1}  &  1.0           \\ 
    \chem{BrMA}     & 7.975\sci{-5}  &   4.032\sci{-5}  & -5.857\sci{3}  \\
    \hline
  \end{tabular}

    \caption{%
      }
    \label{tab:EigData}
  \end{center}
\end{table}


\end{document}